\documentclass[12pts]{article}

\usepackage[displaymath]{lineno}
\usepackage[dvips]{graphicx}
\usepackage[toc,page]{appendix}
\usepackage{graphicx}
\usepackage{amsmath}
\usepackage[varg]{txfonts}

\usepackage{longtable}

\usepackage[usenames]{color}
\usepackage{xspace}
\usepackage{siunitx}
\usepackage{natbib}
\usepackage{filecontents}

\begin{document}
\title{Mesospheric optical signatures of possible lightning on Venus}

\author{
  F. J. P\'erez-Invern\'on$^{1}$,
  A. Luque$^{1}$, 
 F. J. Gordillo-V\'azquez$^{1}$.
  \\
\textit{$^{1}$Instituto de Astrof\'isica de Andaluc\'ia (IAA),} \\
   \textit{CSIC, PO Box 3004, 18080 Granada, Spain.}
\footnote{Correspondence to: fjpi@iaa.es. 
Article published in Journal of Geophysical Research: Space Physics.}
}
\date{}
\maketitle

\begin{abstract}

A self-consistent two-dimensional model is proposed to account for the transient mesospheric nighttime optical emissions associated to possible intra-cloud (IC) lightning occurring in the Venusian troposphere. The model calculates the mesospheric (between 75 km and 120 km in altitude) quasielestrostatic electric field and electron density produced in response to IC lightning activity located beween 40 km and 65 km in the Venusian cloud layer. The optical signatures and the densities of perturbed excited atomic and molecular neutral and ionic species in the mesosphere of Venus are also calcutated using a basic kinetic scheme. The calculations were performed for different IC lightning discharge properties \citep{Krasnopolsky1980/CosmicRes}. We found that the calculated electric fields in the mesosphere of Venus are above breakdown values and that, consequently, visible transient glows (similar to terrestrial Halos produced by lightning) right above the parent IC lightning are predicted. The transient optical emissions result from radiative de-excitation of excited electronic states of N$_2$ (in the ultraviolet, visible and near infrared ranges) and of O($^{1}$S) and of O($^{1}$D) in, respectively, the green (557 nm) and red (630 nm) wavelengths. The predicted transient lightning induced glows from O($^{1}$S) can reach an intensity higher than 167 R and, consequently, be above the detection threshold of the Lightning and Airglow Camera (LAC) instrument aboard the japanese Akatsuki probe orbiting Venus since Dec 2015. However, according to our model, successful observations of transient lightning-induced optical glows could only be possible for sufficiently close (300 km or maximum 1000 km) distances.  
\end{abstract}

\section{Introduction}

The search for lightning activity on Venus dates back to December 1978 when the Soviet Venera 11 and 12 probes descended into the atmosphere of Venus and detected a large number of very low frequency (VLF, 10 - 80 kHz) electromagnetic pulses similar to the radio impulses, also called "spherics", produced by remote lightning \citep{Ksanfomaliti1979/SAL, Ksanfomaliti1980/Natur} on Earth. In addition, also in December 1978, the Orbiter Electric Field Detector (OEFD) of the american Pioneer Venus Orbiter (PVO) sampled the Venus night side at an altitude of about 142-250 km and detected whistler like pulses of waves (in the 100 Hz band) propagating through the Venusian ionosphere \citep{Taylor1979/Science}. However, successive fly-bys of the Venus night side by the Galileo (from a distance of about 4 Venus radii) and Cassini (at approximately 280 km altitude) probes during the 1990s did not show unequivocal high-frequency (0.1 - 16 MHz) electromagnetic signature of the lightning activity that could be taking place in Venus \citep{Gurnett1991/Sci, Gurnett2001/Natur}. More recently,  the magnetometer on board the Venus Express (VEX) probe orbiting Venus (with periapsis in the night side between 250 km and 350 km at the north pole) since April 2006 detected repeated radio bursts (up to 128 Hz) interpreted as whistler mode waves \citep{Russell2007/Natur, Russell2008/JGR, Daniels2012/JGR, Russell2013/GRL}. However, in spite of the Veneras, PVO and VEX impulsive low frequency radio wave detections in Venus, no optical emissions from the night side of Venus have been recorded so far to unambiguously reveal the existence of lightning on Venus. Moreover, no conclusive results have been obtained in the last 20 years from the different attempts to detect lightning activity in Venus using ground-based telescopes \citep{Hansell1995/Icarus, Krasnopolsky2006/Icar, Garcia2013/GRL}.  

On December 2015 the Japanese Venus Climate Orbiter (Akatsuki) was successfully inserted in the Venus orbit opening up new opportunities to directly observe optical emissions from possible lightning occurring on that planet. The Akatsuki probe carries on-board the high-speed Lightning and Airglow Camera (LAC) that could be able to detect possible 777.4 nm lightning emissions and 557.7 nm atomic oxygen nightglow emissions from the Venus nightside \citep{Takahashi2008/SSRv}.

As on Earth, it is possible that lightning on Venus is accompanied by transient luminous events (TLEs) in the mesosphere. Both sprites and halos are generated by lightning-produced quasi-electrostatic electric fields as first proposed by \cite{Wilson1925/PPhSocLon} while elves originate in the lower ionosphere of the Earth due to the propagating lightning electromagnetic pulses \citep{Fukunishi1996/GeoRL}. As suggested by \cite{Yair2009/JGRE}, a possible way to optically detect lightning on Venus would be to search for the transient optical emissions of their induced glows in its upper atmosphere. Such Venusian mesospheric transient optical glows would be generated if sufficiently intense IC lightning flashes take place between the clouds of Venus \citep{Yair2009/JGRE}. In this regard, IC charge moments of about 500 C km have been suggested to produce possible transient optical glows (sprites and/or halos) in Venus at an altitude of around 90 km above ground \citep{Yair2009/JGRE}. 

Recent laboratory experiments \citep{Dubrovin2010/JGR} simulating TLE glows in a 50 mbar (equivalent to 65 km altitude) Venus atmosphere 
(CO$_2$/N$_2$ with a 96.5/3.5 ratio) indicate that the dominant spectroscopic optical signature is in the 300 - 400 nm range or near ultraviolet (NUV) associated to the second positive system (SPS) of N$_2$. An order of magnitude weaker optical emissions in the visible range (400 - 700 nm) were also recorded associated to the first positive system (FPS) of N$_2$ and to several CO transitions. It should be expected, however, that at higher altitudes (lower pressures), the contribution of the FPS optical emissions would become higher (even dominant) due to less efficient quenching as is the case in optical emissions from sprites and halos in the Earth mesosphere \citep{Sentman1995/GeoRL, Hampton1996/GeoRL, Wescott2001/JGR/1, Kanmae2007/GeoRL}.

\section{Model of IC Lightning on Venus}

Following \cite{Yair2009/JGRE} we assume that electrical activity on Venus is mainly due to IC lightning discharges taking place beween or within clouds in the 40 km - 65 km altitude range of the Venusian atmosphere where a layer of sulfuric acid clouds is located. An IC discharge can be considered as an electric dipole ($p$) that induces negative electric charge in the Venusian upper atmospheric layers immediately below the ionosphere. The total electric field in the ionosphere is $E = 0 = E_p + E_S$ where $E_p$ and $E_S$ are the electric fields produced 
by, respectively, the dipole $p$ and a negative screening charge induced by the dipole below the ionosphere. As the IC proceeds, the induced subionospheric negatively charged layer moves down towards lower mesospheric altitudes so that when the original IC discharge is gone, that is, when its equivalent electric dipole is discharged, negative charge and its complementary positive charge remain at mesospheric altitudes (beween 40 km and 65 km) due to the longer Maxwell relaxation times at lower Venusian altitudes where electrical conductivity is smaller than at ionospheric altitudes. The induced negative and positive charge layers at mesospheric altitudes can be imagined as a new emerging electric dipole 
$p'(t) = p(t) - p(0)$ (with $p'(t=0) = 0$ and $p(t=0) = p(0) \neq 0$) that charges up as time progresses and generates an electric field $E_{p(t)'} = E_{p(t)} - E_{p(0)}$. Our IC lightning model here corresponds to the charging electric dipole.

We consider that the center of the IC lightning channel is located around 45 km above ground where the ambient pressure is 1 bar. We also assume that the charge centers are vertically separated by a distance $a$ = 10 km, that is, the IC discharge takes place between vertically aligned clouds at 40 km and 50 km above the surface of Venus. Following \cite{Dubrovin2014/Icarus} we assume that the charge is being accumulated in uniformly charged, non-overlapping identical spheres with a radius of $R$ = 2.5 km \citep{Maggio2009/JGR}. By considering that the total energy released in Venus IC lightning ranges between 8$\times$10${^8}$ and 10$^{10}$ J \citep{Krasnopolsky1980/CosmicRes} and 10$^{11}$ J as an extreme case, the calculation of the electrostatic energy stored by this configuration, 
\begin{linenomath*}
\begin{equation}
U_p = \frac{2Q_T^2}{4 \pi \epsilon_0}\left(\frac{3}{5R} - \frac{1}{2a}\right),
\end{equation}
\end{linenomath*}
allows to estimate that the total dissipated or accumulated charge $Q_T$ is between 15 C and 170 C. Defining the total charge moment change (CMC) as for terrestrial lightning, that is, the product of the transmitted charge and the cloud to ground distance, we have $M = Q_Ta/2$. Thus we have that our model calculations are performed for IC lightning flashes with CMCs between 75 C km and 850 C km. We further assume that the current flowing through the lightning channel follows a bi-exponential function of the form 
\begin{linenomath*}
\begin{equation}
I(t) = I_0 \left(\exp(-t/\tau_1) - \exp(-t/\tau_2)\right),
\end{equation}
\end{linenomath*}
where $\tau_2$ is the rise time of the current wave, and it is typically 10 times faster than the overall durarion of the stroke, represented by $\tau_1$. Considering recent results on the average duration of $\simeq$ 1ms for IC flashes on the Earth \citep{Gaopeng2010/GRL}, we have considered that for a hypothetical 10 km length lightning channel on Venus, the total IC lightning duration will be of the order of $\tau_1$ = 1 ms with a rising time $\tau_2$ $\simeq$ 0.1 $\times$ $\tau_1$ = 0.1 ms in agreement with recordings of IC lightning times on Earth \citep{Rakov2003/ligh.book}. Finally, the accumulated electric charge in our Venus IC lightning model grows as 
\begin{linenomath*}
\begin{equation}
Q(t) = \int_0^t I(t')dt',
\end{equation}
\end{linenomath*}
so that $Q(t=0)=0$ and $Q(t \to \infty)=Q_T$. 

\section{Electric Breakdown in the Nightside Upper Mesosphere of Venus}

Electric breakdown in the Earth upper atmosphere occurs when the reduced electric field ($E/N$, where $E$ is the field strength and $N$ is the ambient number density of neutral species) exceeds the conventional breakdown reduced electric field ($E_k/N$). Due to the decreasing trend of the ambient neutral density with increasing altitude the reduced electric field $E/N$ increases with altitude. The conventional breakdown reduced field $(E_k/N)$ is defined by the competition between two opposing processes in the gas: impact ionization of neutrals by accelerated electrons, and attachment of electrons to certain molecules in the gas (O$_2$ on Earth, CO$_2$ on Venus). According to the classical (no consideration of detachment) electric breakdown mechanism, when the lightning induced electric field exceeds 
$E_k/N$ below the ionosphere then ionization dominates, and electric breakdown can occur. Depending on the values of the parent lightning CMCs different types of Transient Lumiouns Events (TLEs) occur on the Earth mesosphere produced by the quasi-electrostatic and induction electric fields (halos and sprites) and/or the lightning electromagnetic pulse (elves). Our analysis in this paper focuses on the possible existence of halos, visible faint transient glowing regions, in the nightside Venusian mesosphere in response to IC lightning activity. 

\subsection{Ambient Nighttime Electron Density Profile on Venus}

In order to determine the altitude where possible halos can occur in Venus we first need to quantify the ambient nighttime electron density profile  from the cloud deck ($\simeq 60$ km) up to the lower ionosphere ($\simeq 120$ km). However, the available measurements of the Venusian nighttime and daytime electron density at the lowest altitude were recorded by the PVO mission at around 120 km. Below 120 km we have to rely on modeling. According to \cite{Borucki1982/Icarus} and \cite{Marykutty2009/JGR} the nightside ambient electron density in Venus is negligibly small below 60 km but it rapidly increases by three orders of magnitude between 60 km and 75 km. There are no available models for the electron density above 75 km. To calculate the electron density between 75 km and 120 km we have implemented a basic kinetic model considering neutrals \citep{Krasnopolsky2010/Icarus} and ionization profiles from solar extreme ultraviolet (EUV) and X rays and galactic cosmic rays \citep{Nordheim2015/Icarus} controlling the ionization of the atmosphere. The solar and cosmic galactic ray ionizing radiation produce electrons and positive ions such as CO$_2^+$ here and O$^+$ (see discussion below) and lighter (much less abundant) ions like H$^+$ and He$^+$ \citep{Tsang2015/PSC}. The total gas density as a function of the altitude is taken from a Venus reference atmosphere \citep{Valverde2007/PSC} considering CO$_2$/N$_2$ with a 96.5/3.5 ratio and a constant gas temperature of 210 K in the 75 - 120 km altitude range. We also consider the input profiles for the densities of CO \citep{Gilli2010/PSC}, O$_2$ and O \citep{Krasnopolsky2010/Icarus}. 

First we solve a system of 14 stiff balance equations in daytime conditions for a sufficiently long time so that steady-state is achieved at the different altitudes considered. The model output gives the densities of 14 species (electrons, positive and negative ions and some neutrals) (see Table 1) reacting through a set of 40 kinetic reactions (see Table 2). Our daytime model solutions agree with, respectively, available model results \citep{Borucki1982/Icarus, Marykutty2009/JGR} at around 70 km (where solar radiation does not influence atmospheric ionization) and with PVO daytime electron density mearurements at around 120 km. 

We are interested in the Venus nightside electron density profile. However, the solar-antisolar circulation driven by day-night contrast in solar heating occurring above $\simeq$ 110 km of altitude produces strong winds with speeds of $\simeq$ 120 m s$^{-1}$ \citep{Bougher2006/PSS, Lellouch1997/Book}) that can influence the nightside electron density profiles carrying electrons and ions from the dayside and our model can not account for the winds effects. Thus, we stop our electron density simulations at $\simeq$ 105 km where solar ionization is negligible \citep{Nordheim2015/Icarus}. At this altitude we perform linear interpolation up to the PVO nightside electron density recordings at $\simeq$ 120 km. 

\begin{figure}
\includegraphics[width=.55\columnwidth] {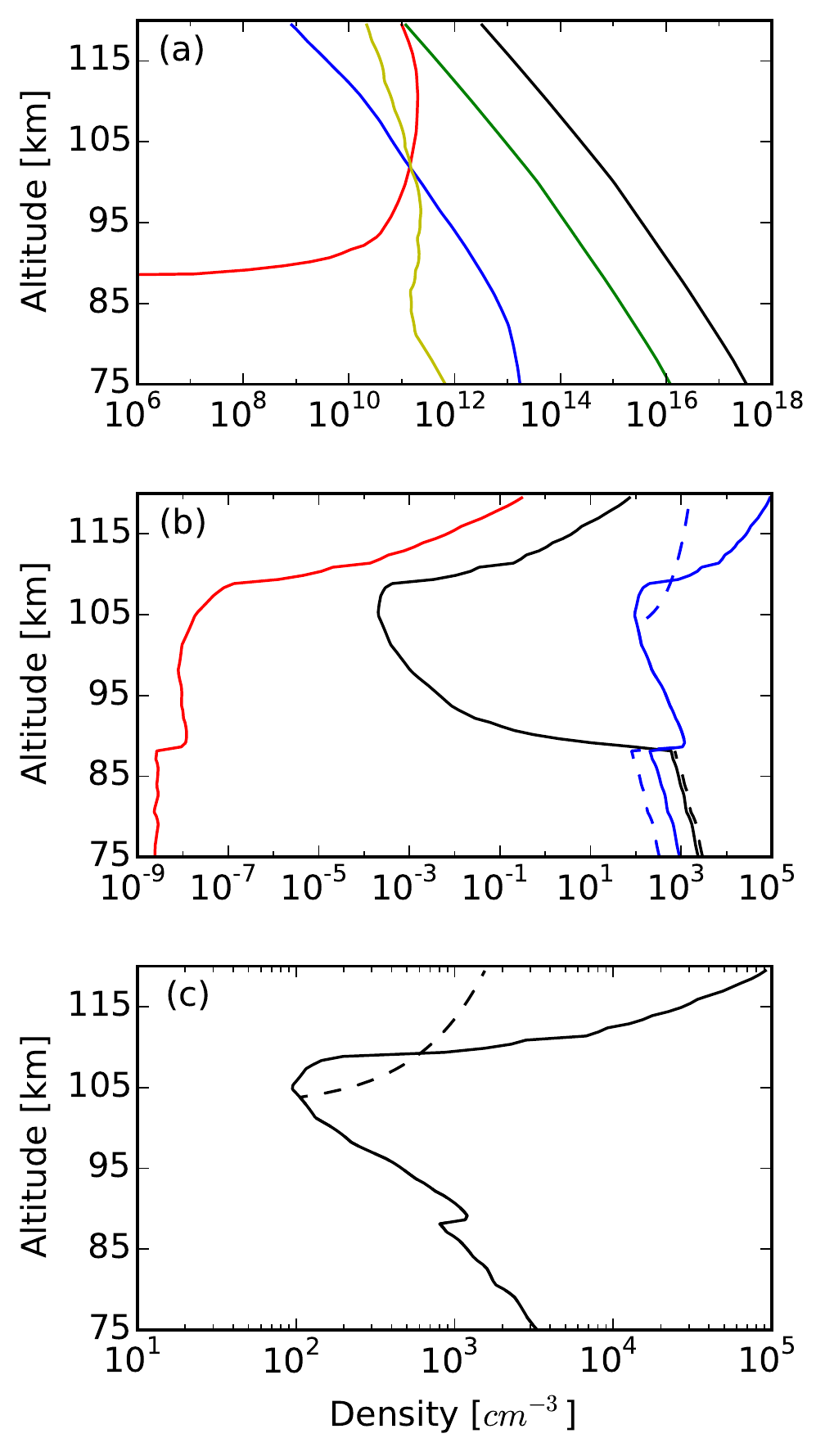}
\centering
\caption{\label{Fig1_Venus_nighttime.pdf} Ambient daytime (solid lines) and nighttime (dashed lines) concentrations of (a) neutrals CO$_2$ (black line), N$_2$ (green line), O$_2$ (blue line), O (red line) and CO (yellow line), (b) positive ions CO$_2$$^{+}$ (black line), O$_2$$^{+}$ (blue line) and O$^{+}$ (red line), and (c) electrons in the mesosphere of Venus as a function of the altitude.}
\end{figure}

The ambient concentrations of the considered neutral species are shown in Figure~\ref{Fig1_Venus_nighttime.pdf} (a). We assume that the concentration of these neutrals remain constant in time. The calculated daytime and nighttime ambient concentrations of positive ions are represented in Figure~\ref{Fig1_Venus_nighttime.pdf} (b). It should be noted that the dominant positive ions are CO$_2$$^{+}$ and O$_2$$^{+}$ (not O$^{+}$) because, according to our model, once O$^+$ is produced by the photoionization processes considered \citep{Nordheim2015/Icarus} it is rapidly converted into O$_2$$^{+}$ through the reaction O$^{+}$ + CO$_2$ $\rightarrow$ CO + O$_2$$^{+}$. In the altitude range (70 - 125 km) investigated, where the concentration of CO$_2$ prevails over that of O, our kinetic model predicts that O$_2$$^{+}$ dominates over O$^{+}$ in agreement with nighttime PVO observations and the Venus International Reference Atmosphere (VIRA) \citep{Bauer/ASR}.

\begin{figure}
\includegraphics[width=.9\columnwidth] {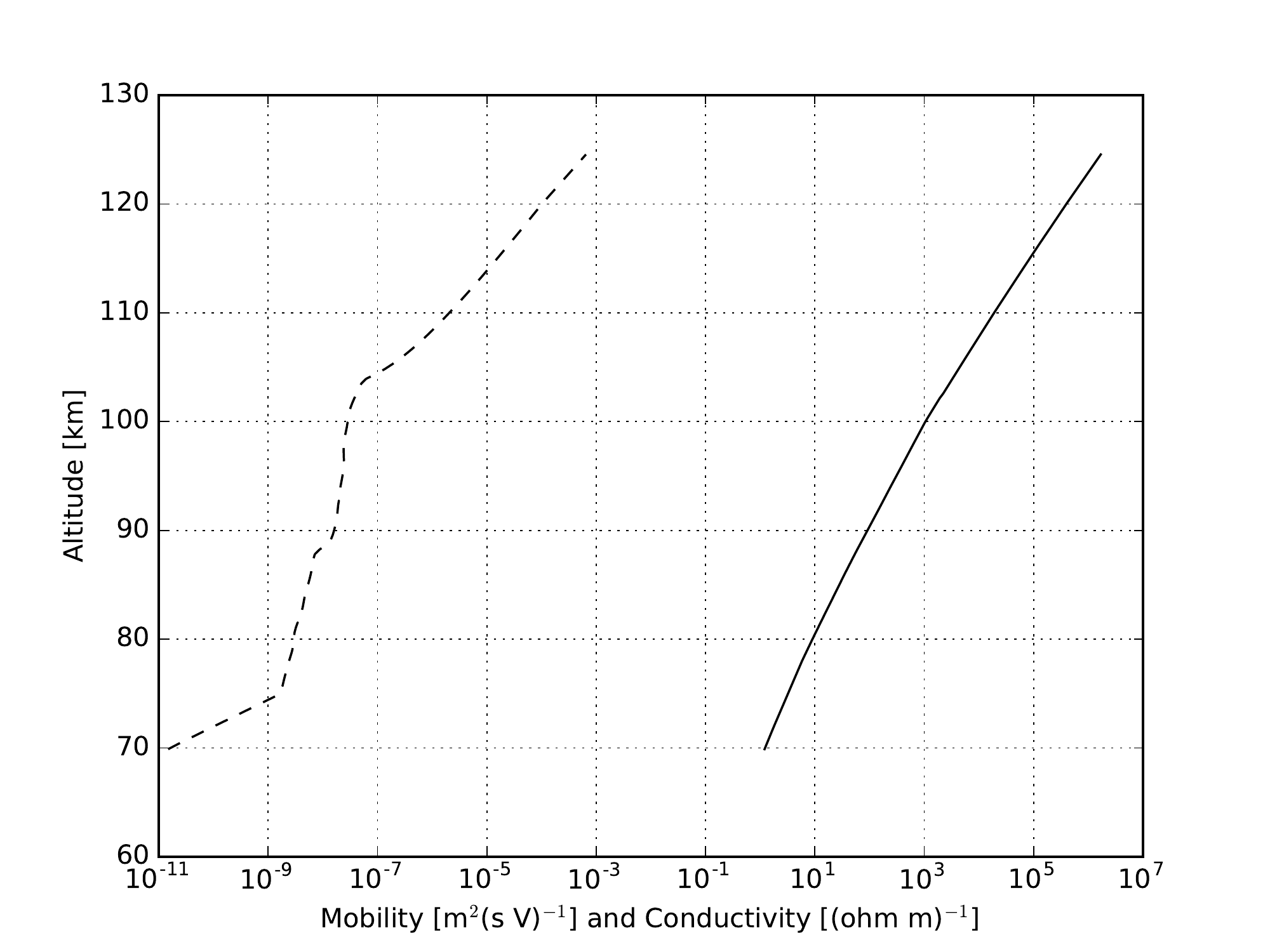}
\caption{\label{Fig2_Venus_nighttime.pdf} Mobility (solid line) and conductivity (dashed line) ($\sigma$ = $e$$\mu_e$$N_e$) of nighttime electrons in the mesosphere of Venus assuming $E/N$ = 0 (ambient conditions).}
\end{figure}

Figure~\ref{Fig1_Venus_nighttime.pdf} (c) shows the ambient daytime and nighttime electron density profiles in the mesosphere of Venus resulting from our calculations. The maximum electron density in the nighttime profile between 75 km and 120 km is located at around 75 km influenced by the proximity to the maximum of the galactic cosmic ray ionization taking place at 63 km \citep{Nordheim2015/Icarus}. The mobility and conductivity ($\sigma$ = $e$$\mu_e$$N_e$) of ambient nighttime electrons are shown in Figure~\ref{Fig2_Venus_nighttime.pdf}. The mobility of electrons was obtained by calculating the electron energy distribution function \citep{Hagelaar2005/PSST} in a gas mixture with the proportions of the Venus atmosphere and assuming a negligible ($\simeq$ 0 Td) ambient reduced electric field. The mobility is scaled up for the different pressures (altitudes) considered. The calculated values of the mobility and conductivity of ambient nighttime electrons show a growing trend with increasing altitudes and their lower bound (70 - 80 km) values agree well with previous calculations by \cite{Marykutty2009/JGR} in the absence of cloud particles and available only up to altitudes of 70 km.

\subsection{Electric Breakdown on Venus}

Kinetic mechanisms such as electron-impact ionization of CO$_2$ and N$_2$ (e + CO$_2$ / N$_2$ $\to$ CO$_2$$^+$ / N$_2$$^+$ + 2e) and electron-driven dissociative attachment of CO$_2$ (e + CO$_2$ $\to$ CO$^{-}$ + O) are both efficient processes in the atmosphere of Venus. The conventional breakdown reduced electric field ($E_k/N$) is obtained when the effective ionization frequency $\nu_{i, eff}$ = $\nu_{i}$ - $\nu_{att}$ is equal to zero, that is, when the ionization ($\nu_{i}$) and attachment ($\nu_{att}$) frequencies become equal. Our calculations indicate that  $E_k/N$ $\simeq$ 74 Td in the atmosphere of Venus.

\begin{figure}
\includegraphics[width=.9\columnwidth] {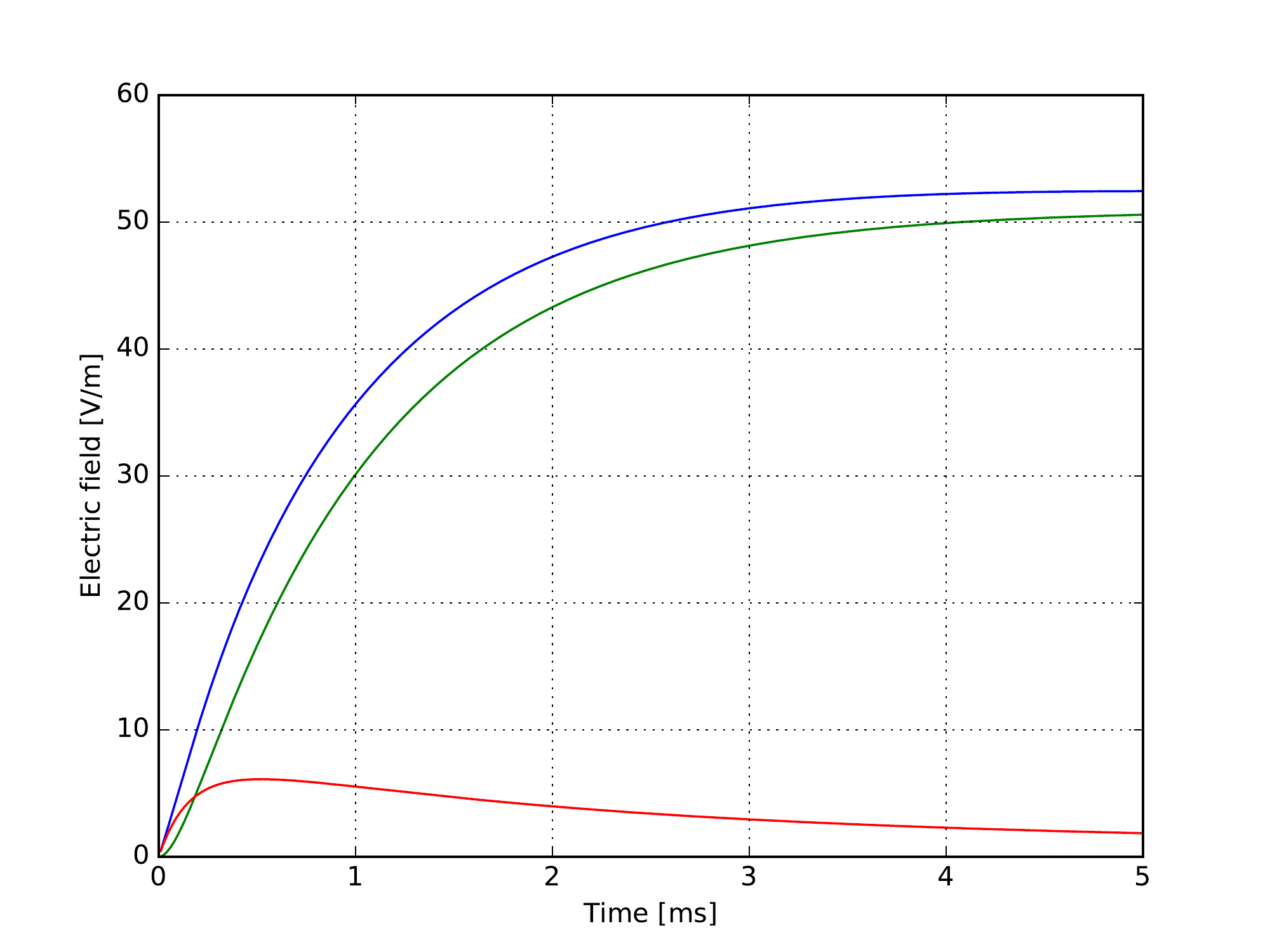}
\caption{\label{Fig3_Venus_nighttime.png} Time evolution of the applied electric field (blue), static field (green) and induction field (red)  components at 100 km due to an in-cloud discharge with a released energy of 10$^{10}$ J calculated according to Eq.  (\ref{ec3}).}
\end{figure}

\begin{figure}
\includegraphics[width=.9\columnwidth] {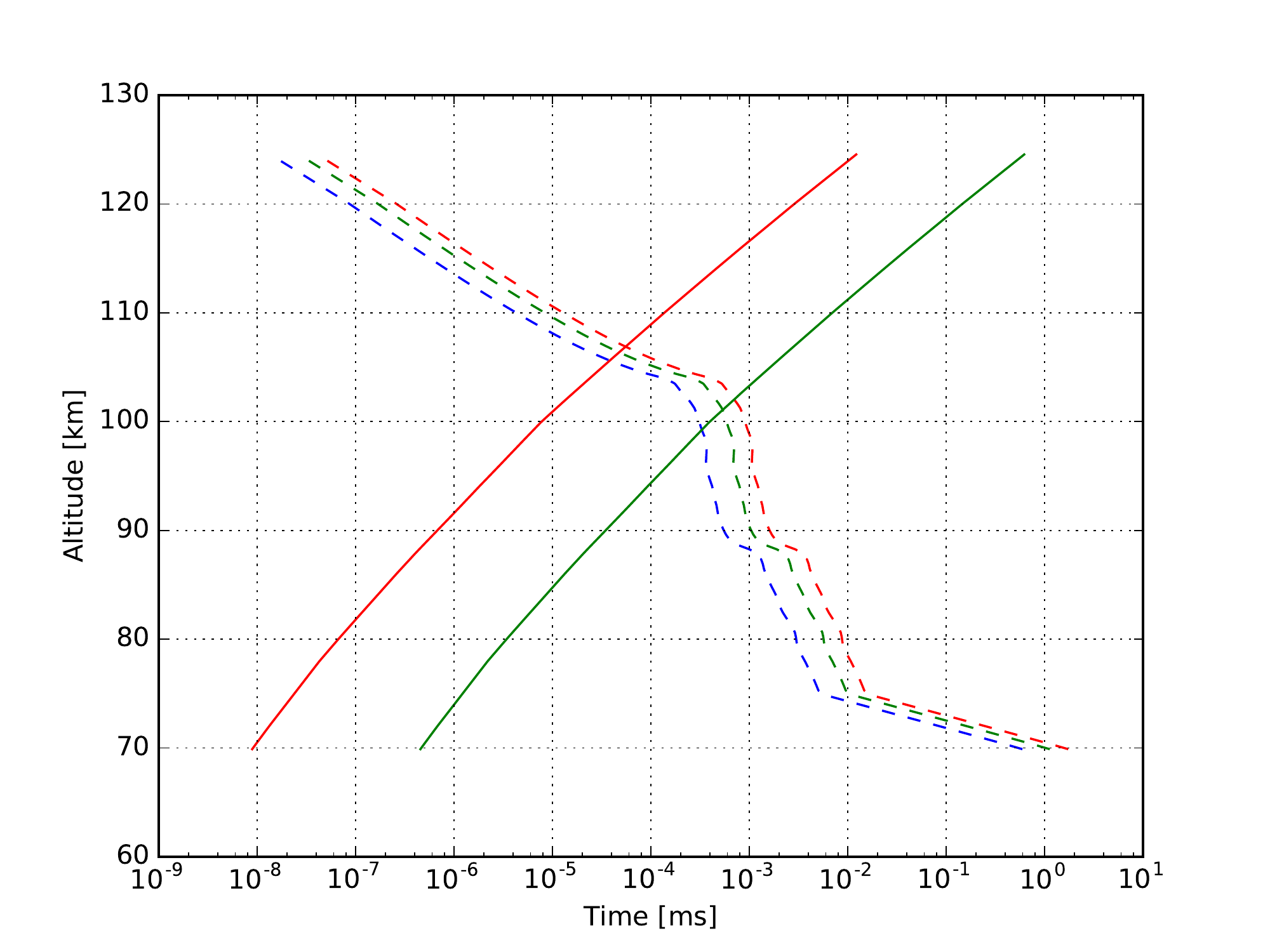}
\caption{\label{Fig4_Venus_nighttime.pdf}  Maxwell relaxation times for $E/N$ = 0  (ambient conditions) (dashed blue line), $E/N$ = 80 Td  (dashed green line) and $E/N$ = 2 $E_k/N$ (dashed red line). Effective ionization times for $E/N$ = 80 Td (solid green line) and $E/N$ = 2 $E_k/N$ (solid red line).}
\end{figure}

In order to determine a realistic response of the Venus atmosphere to a possible lightning-produced electric field in the Venusian mesosphere we need to consider different but interconnected time scales related to the electric field, $\tau_E$, that is, its rise time ($\tau_2$) and/or decay times ($\tau_1$); the Maxwell relaxation time, $\tau_M = \epsilon_0 / \sigma$, determined by the local electron conductivity ($\sigma$) with $\epsilon_0$ being the electric permittivity of vacuum, and the effective ionization time, $\tau_{i, eff} = 1 / \nu_{i, eff}$. The screening of an applied electric field can only be avoided if its time scale (for rise and/or decay) are shorter than the Maxwell time. 
We can calculate the time evolution of the electric field in a given altitude z as the sum of the quasi-electrostatic field produced by a dipole and the induction field proportional to the discharging current:
\begin{linenomath*}
\begin{equation}
 E(z,t) = \frac{1}{\pi\epsilon_0} \left(\frac{1}{(z-z_p)^{3}}M(t) + \frac{1}{c(z-z_p)^{2}}\frac{d}{dt}M(t)\right)
 \label{ec1}
\end{equation}
\end{linenomath*}
where $M(t)$ is the charge moment change and $z_p$ is the dipole altitude.
As can be seen in Figure~\ref{Fig3_Venus_nighttime.png} the induction electric field (proportional to the current intensity and controlled by the rise time ($\tau_2$)) is considerably smaller than the quasi-electrostatic (QES) electric field beyond $\tau_2$ = 0.1 ms. Therefore, screening of the QES electric field is avoided in regions where $\tau_1$ $\leq$ $\tau_M$. In addition to avoid electric screening, the applied electric field needs to have enough time to ionize the atmosphere which also requires that $\tau_{i, eff}$ $\leq$ $\tau_M$. We see in Figure~\ref{Fig4_Venus_nighttime.pdf} the Maxwell relaxation time and the effective ionization time for the nighttime electron density profile in the mesosphere of Venus calculated for two values of the reduced electric field, close to breakdown ($E/N$ = 80 Td $\simeq$ 1.1 $E_k/N$) and for $E/N$ = 148 Td = 2 $E_k/N$.  The Maxwell relaxation time for ambient conditions ($E/N$ = 0 Td) is also shown in Figure~\ref{Fig4_Venus_nighttime.pdf}. According to 
Figure~\ref{Fig4_Venus_nighttime.pdf} electric breakdown due to QES fields with amplitude $E/N$ $\simeq$ $E_k/N$ will occur at  altitudes between $\simeq$ 95 km and $\simeq$ 100 km when $\tau_1$ $\leq$ 0.9 ms and $\leq$ 0.8 ms, respectively.

\section{Model of Electrical Discharges in the Venusian Upper Mesosphere}

We have developed a 2D model in cylindrical coordinates to account for the transient electrical discharges in the mesosphere of Venus due to possible IC lightning activity within the cloud deck. The model aims to quantify the lightning-induced transient electric fields in the Venusian mesosphere, the mesospheric chemical influence of IC lightning and the excited chemical species responsible for possible transient optical emissions (glows) that could be tracked from orbit to optically confirm the existence of lightning on Venus.

\subsection{Model equations and numerical implementation}

The 2D model solves the following set of coupled equations: 
\begin{linenomath*}
\begin{equation}
   \nabla^2 \phi = - \frac{\rho}{\epsilon_0}
 \label{ec2}
\end{equation}
\end{linenomath*}
\begin{linenomath*}
\begin{equation}
   \frac{\partial N_e}{\partial t} + \nabla \cdot \mathbf{J}_e = P_e - L_e
 \label{ec3}
\end{equation}
\end{linenomath*}
and continuity equations for each of the $i = 1,\dots 27$ neutral and ionic species considered
\begin{linenomath*}
\begin{equation}
   \frac{\partial N_i}{\partial t} = P_i - L_i.
 \label{ec4}
\end{equation}
\end{linenomath*}
where $P_{e,i}$ and $L_{e,i}$ indicate, respectively, 
the production and loss of electrons ($e$) or heavy species $i$.

The magnitudes $\phi$, $\epsilon$ and $\rho$ in Poisson's equation (\ref{ec2}) stand for, respectively, the electric potential, the permittivity of vacuum and the charge density in the two domains of integration considered below 70 km and above 70 km where chemical kinetics is solved. Above 70 km, $\rho = q_e (N_i^+ - N_e - N_i^-)$ where $q_e$ is the elementary charge and $N_i^+$, $N_e$ and $N_i^-$ indicate the calculated concentrations of positive ions, electrons and negative ions. Below 70 km, $\rho = Q^+/V - Q^-/V$, that is, the positive and negative charge densities (within a cylindrical volume $V = 2 \pi R^3$) of an electric dipole representing an IC lightning. 

The magnitude $J_e$ in the advection-diffusion equation (\ref{ec3}) for electrons stand for the axial 

\begin{subequations}
\begin{linenomath*}
  \begin{equation}
    J_{e, z} = -D_e \frac{\partial N_{e, z}}{\partial z} - \mu_{e, z} E_z N_{e, z},
  \end{equation}
\end{linenomath*}
and radial 
\begin{equation}
  J_{e, r} = -D_e \frac{1}{r} \frac{\partial N_{e, r}}{r\partial r} - \mu_{e, r} E_r N_{e, r}
\end{equation}
\end{subequations}

flows of electrons produced by the induced electric field that try to screen it out. The electron diffusion ($D_e$) and mobility ($\mu_e$) are computed offline  by solving the steady-state Boltzmann equation \citep{Hagelaar2005/PSST} as a function of the reduced electric field for a Venusian gas mixture. We found that in the $E/N$ range of interest (0 - 400 Td) and for the length scale of a few kilometers that we consider, the diffusion of electrons is negligible so that we can assume 
\begin{subequations}
\begin{linenomath*}
\begin{align}
  J_{e, z} & \simeq - \mu_{e, z} E_z N_{e, z} \\
  J_{e, r} & \simeq - \mu_{e, r} E_z N_{e, r}.
\end{align}
\end{linenomath*}
\end{subequations}

In order to solve the system of equations (\ref{ec2})-(\ref{ec4}) we have used a fixed cylindrical grid with spatial and temporal resolutions given by, respectively, $\Delta z$ = 167 m, $\Delta r$ = 250 m and $\Delta t$ = 10$^{-8}$ s . We have discretized the axial and radial advective terms of equation (\ref{ec3}) by using a Koren limiter function \citep{Montijn2006/JCoPh} so that possible undesired numerical diffusion and oscillations due to the abrupt axial changes in the electron density are avoided. The boundary conditions for the Poisson equation (\ref{ec2}) are of the Neumann type at $r$ = 250 km, undetermined in the axis of symmetry and of the Dirichlet type in the upper (ionoshere) and lower (ground) boundaries assuming that both of them behave as perfect electric conductors (PEC). On the other hand, Neumann type boundary conditions are used for solving equations (\ref{ec3})-(\ref{ec4}) in the region where plasma kinetics is considered, that is, 70 km $\leq$ z $\leq$ 125 km and 0 km $\leq$ r $\leq$ 200 km. The temporal dependence of the continuity equations for electrons and for species $i$ (equations (\ref{ec3}) and (\ref{ec4})) is treated by using an explicit runge-kutta method of order 5 with step size control based on the Dormand and Prince algorithm \citep{DormandPrince1980/JAM}. We use the solver FISHPACK \citep{Fishpack} in cylindrical coordinates to solve the 2D Poisson equation (\ref{ec2}) with a vertical IC discharge located in r = 0 km.

\subsection{Reduced Kinetic Scheme for the mesosphere of Venus}

We have implemented a reduced kinetic scheme for the upper atmosphere of Venus in order to investigate how it changes under the influence of lightning-produced electric fields. The scheme considers a number of neutral species (CO$_2$, N$_2$, CO, O and O$_2$) with known concentrations that do not change in time. The kinetic scheme is applied between 70 km and 125 km of altitude where perturbations due to underlying suspected lightning activity is possible. The initial conditions of the 2D discharge model are the ambient ($E/N$ $\simeq$ 0 Td) altitude-dependent profiles for electrons and key positives ions CO$_2$$^{+}$ and O$_2$$^{+}$ (N$_2$$^+$ is rapidly converted into O$_2$$^{+}$) obtained from the equilibrium model described in section 3.1

We include as additional inputs the ambient ($E/N$ $\simeq$ 0 Td) altitude-dependent profiles of eight vibrational levels of CO$_2$ ((00$^0$1),  (10$^0$0), (01$^1$0), (02$^0$2), (02$^2$2), (03$^1$0), (03$^3$0) and (11$^1$0)) previously calculated under steady-state nonlocal thermodynamic equilibrium (N-LTE) \citep{Valverde2007/PSC}.  The reduced kinetic scheme of our discharge model includes electron-impact excitation of CO$_2$, vibrarional-vibrational (VV), vibrational-translational (VT) processes and radiative deexcitation producing IR optical emissions. However, because of the differences between our reduced kinetic sheme and the N-LTE kinetics used to calculate ambient CO$_2$(v) concentrations, our discharge model produces $dN_{CO_2(v)}(E/N=0)/dt$  = $A(v)$ $\neq$ 0. In order to have $dN_{CO_2(v)}(E/N=0)/dt$ = 0 we re-formulate the continuity equations for the considered CO$_2$(v) by subtracting $A(v)$ to each of them.

The output of our 2D electric discharge for the mesosphere of Venus provides space-time dependent concentrations of electrons and 27 species (see Table 3) interacting through 80 kinetic reactions (see Table 4 and 5) that include heavy-heavy reactions (such as, among others, quenching and charge transfer), radiative de-excitation and electron-driven processes (with electric field dependent rates) such as ionization, attachment, detachment, recombination and electronic and vibrational excitation of atomic and molecular species. Among the species considered the excited O($^1$S) atoms can decay radiatively emitting the green (557.7 nm) line detected on the nightside (atomic oxygen nightglow) of Venus with considerable variations in the recorded intensity brightnesses that changes between less than 10 R (with periods when the green line is absent) and 
167 R \citep{Slanger2006/Icarus, Gray2014/Icarus}. 

In addition to O($^1$S) and O($^1$D) atoms, we have also considered several electronic excited states of N$_2$ and vibrational levels of CO$_2$ that can be produced through electron impact collisions. In particular, we have included possible optical emissions from N$_2$(B$^3\Pi_g$), N$_2$(C$^3\Pi_u$) and N$_2$(a$^1\Pi_g$) involved in, respectively, the first positive system (FPS), second positive system (SPS) and the Lyman Birge Hopfield (LBH) system of N$_2$. Optical emissions from the LBH, SPS and FPS systems are in the ultravioled (150 - 250 nm), near ultraviolet-blue (250 nm - 450 nm) and in the visible-near infrared (500 nm - 1.5 $\mu$m).  We have also included infrared emissions corresponding to the transition N$_2$($W^{3}\Delta_u$) $\to$ N$_2$($B^{3}\Pi_g$).

\section{Electric Field and Electron Density}

The existence of possible lightning activity on Venus can induce the emergence of transient quasi-electrostatic electric fields in the mesosphere of Venus. Figures 5 and 6 show different snapshots of the reduced electric fields $(E/N)$ and electron density under different lightning discharges located at r = 0 km. We can see that for lightning energies above 10$^{10}$ J, around 0.25 - 0.75 ms after lightning is produced, the electric field takes values above the Venusian breakdown field producing regions of significant ionization where the concentration of electrons increases causing a rapid screening in the electric field. As expected, the maximum value of the reduced electric field is reached in the region where the ambient electron density takes its minimum value, at an altitude of 100 - 105 km (see Figure 1). 

We have also explored the influence of changing different lightning characteristics on the reduced electric field of the lightning-induced glowing halo generated in the Venusian upper atmosphere. For instance, shorter lightning lengths produce lower reduced electric fields. Moreover, the radii of the spheres that represent the cloud charges also influence the total reduced electric field since, according to \cite{Maggio2009/JGR}, for a given a total released lightning energy, an increase in the radii of the spheres produces larger amounts of total transferred charge.

\section{Evolution of vibrational levels of CO$_2$}

Figure 7 shows the time evolution of the CO$_2$(001) density after the beginning of the IC discharge. According to our model, the only vibrational level of carbon dioxide that locally increases due to the action of the reduced electric field of the glowing halo is the 001. As we clearly see in Figure 7(c) the CO$_2$(001) density growth is weak and confined to a region where the total electric field reaches values close to breakdown. The radiative de-excitation of CO$_2$(001) takes place $\simeq$ 2 ms after being excited and emits light in the near infrared (NIR) range at 4.26 $\mu$m.

\section{Transient Optical Fingerprints}

The optical emission brightness (EB) at a given time in Rayleighs (R) is calculated from the species densities through the classical expression \citep{Hunten1956/JATP}:
\begin{linenomath*}
\begin{equation}
 EB(t) = 10^{-6}\int_{L}V(l,t) \ dl
 \label{ec5}
\end{equation}
\end{linenomath*}
where $V(l,t)$ is the volume emission rate of each species (in photons cm$^{-3}$s$^{-1}$) at a given time, and the integral is taken along the line of sight through the emission volume over an effective column lenght $L$.
The total number of photons per second emitted at a given time ($E_{h\nu}(t)$) are calculated integrating $V(l,t)$ over the whole volume domain taking into account the cylindrical simmetry:
\begin{linenomath*}
\begin{equation}
 E_{h\nu}(t) = \int_{V}V(l,t) \ dV
 \label{ec6}
\end{equation}
\end{linenomath*}

\subsection{Atomic Oxygen Optical Emissions}
Figure 8 shows the time dependent emission brightness of the O($^{1}$S) $\to$ O($^{1}$D) optical transition (at 557.7 nm) from the center to the outer region of the transient glowing halo and for three snapshots (0.5 ms, 0.75 ms and 1 ms) after the beginning of the IC discharge. We can see that the maximum brightness occurs in the center (r = 0 km) of the halo with the strongest optical emissions being due to IC discharges with total energies above 2 $\times$ 10$^{10}$ J (middle panel) producing relatively high reduced electric fields.

In the mesosphere of Venus, where the total density is not high, the radiative decay of O($^{1}$S) (see reaction (65) in Table 4) dominates over collisional deactivation (reactions (61) to (64)), causing an increase in the concentration of O($^{1}$D) in a timescale of several hundreds of miliseconds. After formation, O($^{1}$D) will decay radiatively emitting red (630 nm) photons in a much longer timescale ($\simeq$ 200 s) than that of the glowing halo. 

It is finally interesting to note that the reported brightnesses of the O($^{1}$S) nightglow in Venus is latitude-dependent and their values change between few Rayleighs and a maximum of 167 Rayleighs \citep{Slanger2006/Icarus, Gray2014/Icarus}. 

\subsection{Molecular Nitrogen Optical Emissions}

Molecular nitrogen is excited by electron impact collisions resulting in electronically excited states including vibrational excitation. In our model, we calculate the concentrations of electronically excited states averaged over all their vibrational levels. All the considered electronically excited states N$_2$(B$^3\Pi_g$ (all v$^{\prime}$)), N$_2$(C$^3\Pi_u$ (all v$^{\prime}$)),  N$_2$($W^{3}\Delta_u$ (all v$^{\prime}$)) and N$_2$(a$^1\Pi_g$ (all v$^{\prime}$)) emit photons due to their radiative decays. The wavelengths of these emissions depend on both the initial (upper, v$^{\prime}$) and the final (lower, v$^{\prime\prime}$) vibrational excitation level of the involved upper and lower N$_2$ electronic states. According to Figure 9.1. in \cite{Capitelli/Book}, radiative decay constants for reactions (70), (71) and (72) shown in Table 4 are approximatelly the same for all v$^{\prime}$, so we have used mean radiative decay constants in our model. However, radiative constants for reactions (72)-(74) in Table 4 depend critically on the chosen initial vibrational level of excitation. Therefore, for the triplet W$^3$$\Delta_u$ and singlet a$^1$$\Pi_g$ electronic states we have only used radiative decay constants from the lowest (most populated) vibrational level of the upper electronic state v$^{\prime}$ = 0 (see reactions 70-75 in Table 4).
All the possible optical emissions in the upper atmosphere of Venus caused by possible IC discharges are summarized in Table 6. 

Figures 9-11 show the emission brightnesses (in R) along the radius of the glowing halo due to the radiative decays of N$_2$(B$^3\Pi_g$), N$_2$(C$^3\Pi_u$),  N$_2$($W^{3}\Delta_u$) and N$_2$(a$^1\Pi_g$) for three snapshots (0.5 ms, 0.75 ms and 1 ms) after the beginning of the IC discharge. We can see in Figures 9-11 that the halo radius reaches its maximum value around 40 km and that it mainly depends on the charge moment change of the discharge. The strongest emissions are produced in the optical range associated to the FPS and SPS of molecular nitrogen, while near-infrared and ultraviolet emissions are also possible due to N$_2$($W^{3}\Delta_u$) and N$_2$(a$^1\Pi_g$) states. 

It is worth mentioning that, in order to avoid numerical instabilities, we stopped our simulations at 1 milisecond. Therefore,  since we are only solving our model up to 1 ms, we are somehow underestimating the total brightness and total number of photons emitted per second by the glowing halo. However, as shown in Figure 5, there are thin and very localized regions (see the 10$^{11}$ J case in Figure 5) with high (well above breakdown) reduced electric fields that could be emitting during longer times (tens of miliseconds).

\section{Feasibility of Detecting Venus Transient Optical Emissions from Orbiters}

The lightning induced transient optical emissions predicted by our model are produced over the Venusian cloud deck. We discuss in this section 
the  feasibility of detecting the emitted photons from orbiters around Venus. In this regard, the Akatsuki probe is presently orbitting Venus at distances from periapsis (varying between 300 and 4000 km) to apoapsis at 370000 km \citep{Gibney/Nat} (see Figure 12) and it is equipped with the Lightning and Airglow Camera (LAC) that could be able to detect the transient green (557.7 nm) line produced by the radiative decay of O($^{1}$S) \citep{Takahashi2008/SSRv}. About this possibility, it is worth mentioning that the existence of lightning induced transient airglow enhancements in the Earth was first detected from the space shuttle. On October 1990, the shuttle Discovery (on mission STS-41) in orbit at 296 km to 327 km altitude recorded (in the 360 nm - 720 nm range) a sequence of sudden brightenings at the altitude of the airglow layer in coincidence with lightning flashes \citep{Boeck1992/GRL}. The region of enhanced luminosity was 10 to 20 km thick and around 500 km in diameter with a brightness of about twice that of the background airglow \cite{Boeck1992/GRL}. We now know that what space Shuttles recorded were the fast ($\ll$ 1 ms) transient optical emissions of the first and second positive systems of N$_2$ excited by elves \citep{Fukunishi1996/GeoRL}, \citep{Israelevich2004/GeoRL}. Halos, due to the lightning-induced quasielectrostatic electric fields, have  characteristic scales of $\simeq$ 100 km (shorter than elves) but last longer, about ten times more $\simeq$ 1 ms.  

In addition to LAC, the Akatsuki spacecraft also carries cameras that cover the UV range (258 and 360 nm) and the near-IR (1.0, 1.7, 2.0 and 2.3 $\mu$m) \citep{Takahashi2008/SSRv} covering the spectral range of optical emissions associated to the electronic states N$_2$(B$^3\Pi_g$), N$_2$(C$^3\Pi_u$), N$_2$($W^{3}\Delta_u$) and N$_2$(a$^1\Pi_g$). In this regard, it is important to note that the transient lightning-induced NIR optical emissions in the upper atmosphere of Venus corresponding to the strong (high radiative decay constants) transitions N$_2$($W^{3}\Delta_u($v$^{\prime}$ = 3, 4)) $\to$ N$_2$(B$^3\Pi_g$(v$^{\prime\prime}$ = 0)) centered around, respectively, 2.250 $\mu$m and 1.709 $\mu$m and N$_2$(a$^1\Pi_g$(v$^{\prime}$ = 2, 3)) $\to$ N$_2$($a^{\prime1}\Sigma_{u}^{-}$ (v$^{\prime\prime}=$ 0)) centered around, respectively, 2.214 $\mu$m and 1.632 $\mu$m are within the spectral range of the NIR-IR2 camera of Akatsuki. However, vibrational kinetics is not included in the present version of our model and, consequently,  we can not precisely compute the concentrations of the vibrational levels underlying the above mentioned NIR transitions.

Figures 13-16 show the time-dependent total number of photons emitted per second from the entire transient glowing halo calculated using equation \ref{ec6}. The shown optical emissions per second exhibit a sudden increase (up to around 0.4 ms) when the reduced electric field reaches breakdown producing a strong rise in the densities of excited species.
 
If we numerically integrate in time (up to 1 milisecond) the total number of photons emitted per second we obtain the total number of emitted photons up to the instant when the number of photons emitted per second do not grow any further. From this time on, we use the calculated number density of each emitting species at 1 milisecond to analytically integrate it in time and derive the total number of emitted photon from 1 milisecond to infinity. The sum of both (numerical and analytical) integrals provides the total number of emitted photons given in Table 7 through Table 9 for each of the considered emitting species. Afterwards, we can calculate the fraction of emitted photons able to reach an hypothetical detector with a diameter of 25 mm at a distance $L$ from the halo, located at around 100 km of altitude:

\begin{linenomath*}
\begin{equation}
  \frac{R_{h\nu}}{E_{h\nu}} = \frac{\pi r^{2}}{4 \pi L^{2}},
 \label{ec7}
\end{equation}
\end{linenomath*}
where $R_{h\nu}$ is the total number of photons that reach the camera,  $E_{h\nu}$ is the number of photons emitted by the entire halo, and $r$ is the radius of the camera aperture.

Tables 7-12 show the total number of emitted photons (Tables 7, 9, 11) and the calculated total number of photons received (Tables 8, 10, 12) by a 25 mm diameter camera sensor at a distance of 300 km and 1000 km from a possible Venusian glowing halo. The LAC airglow observation mode uses a section of 1 x 8 pixels of the full (8 x 8 pixels) avalanche photodiode (APD) detector  \citep{Takahashi2008/SSRv}. Therefore, we can conclude that the total amount of photons that reaches the airglow detector section from the transient glowing discharge is 1/8 of the total number of photons received by the full detector shown in tables 7-12.

Figure 8 shows that, for highly energetic IC discharges, the halo brightness in the green line (557.7 nm) is larger than the maximum natural atomic oxygen nightglow emission of 167 R \citep{Slanger2006/Icarus, Gray2014/Icarus}. This means that a camera with sufficiently high temporal resolution and sufficiently close could detect the predicted 557.7 nm transient increase above its ambient nightglow value.

Optical emissions due to the increase in the density of CO$_2$(001) might not be detectable from space due to the strong ambient emission in the 4.26 $\mu$m spectral region \citep{Valverde2007/PSC}. However, we speculate that a probe located on the Venus surface (or close to it) could observe this weak and local transient emission increase corresponding to the 4.26 $\mu$m vibrational transition of CO$_2$.

\section{Conclusions}

The existence of lightning on Venus is still controvesial, as there is no confirmed optical evidence.  Our aim in this paper was to develop a two dimensonal model able to predict and quantify possible mesospheric transient optical emissions caused by possible IC lightning activity on Venus. In order to do so, our discharge model solves a set of differential equations in the space-time domain. Firstly, we relax the system of equations by integrating it for relatively long times in order to obtain an ambient electron density equilibrium profile consistent with available observations and cosmic ray ionization rates. Then we obtained that the reduced electric field that produces electric breakdown of the Venusian atmosphere is 74 Td. We found that the mesospheric quasi-electrostatic electric fields induced by possible IC lightning discharges could only produce detectable transient upper atmospheric glowing halos when the minimum total energy released by Venusian IC lightnings is 2 $\times$ 10$^{10}$ J (orbiter at a distance of 300 km) or 10$^{11}$ J (orbiter at a distance of 1000 km).

According to our model predictions, the strongest mesospheric transient optical emissions are due (in order of importance) to radiative de-excitation of N$_2$(B$^3\Pi_g$), O($^{1}S$) and N$_2$(C$^3\Pi_u$), while other infrared and ultraviolet emissions are produced by radiative decay of N$_2$($W^{3}\Delta_u$) and N$_2$(a$^1\Pi_g$). In particular, our model predicts that the 557.7 nm mesospheric transient optical emissions of O($^{1}S$) could be detectable by orbiters when induced by IC discharges with minimum total energies of 2 $\times$ 10$^{10}$ J. In the latter cases, the transient 557.7 nm emission brightness would be larger than the ambient Venusian atomic oxygen nightglow. Moreover, lightning-induced transient optical emissions from the O($^1$D) radiative decay in the 630 nm line are also expected though for much longer (minutes) time scales.

As we have seen, highly energetic IC lightning discharges could trigger transient optical emissions in the mesosphere of Venus that could be detectable by sufficiently close orbiters. High temporal resolution optical observations of the green line (557.7 nm) could be able to detect a rapid increase-decrease in the optical 557.7 nm emission signal due to underlying lightning activity. The transient 557.7 nm emission signal can be brighter than that of the background green nightglow.

Future observations of transient optical, ultraviolet and infrared emissions associated to the radiative decays of N$_2$ electronically excited species in the mesosphere of Venus could be a feasible way to finally conclude on the existence of lightning on Venus.

\section*{Acknowledgement}
This work was supported by the Spanish Ministry of Science and Innovation, MINECO under projects ESP2013-48032-C5-5-R,  FIS2014-61774-EXP and ESP2015-69909-C5-2-R and by the EU through the FEDER program. FJPI acknowledges a MINECO predoctoral contract, code BES-2014-069567. AL acknowledges support by a Ram{\'o}n y Cajal contract, code RYC-2011-07801. All data used in this paper are directly available after a request is made to authors F.J.P.I (fjpi@iaa.es), A.L (aluque@iaa.es) or F.J.G.V (vazquez@iaa.es).

\newcommand{\jcp}{J. Chem. Phys. } 
\newcommand{\ssr}{Space Sci. Rev.} 
\newcommand{\planss}{Plan. Spac. Sci.} 
\newcommand{\pre}{Phys. Rev. E} 
\newcommand{\nat}{Nature} 
\newcommand{\icarus}{Icarus} 
\newcommand{\ndash}{-} 
\newcommand{\jgr}{J. Geophys. Res.}

\begin{figure}
\includegraphics[width=.9\columnwidth] {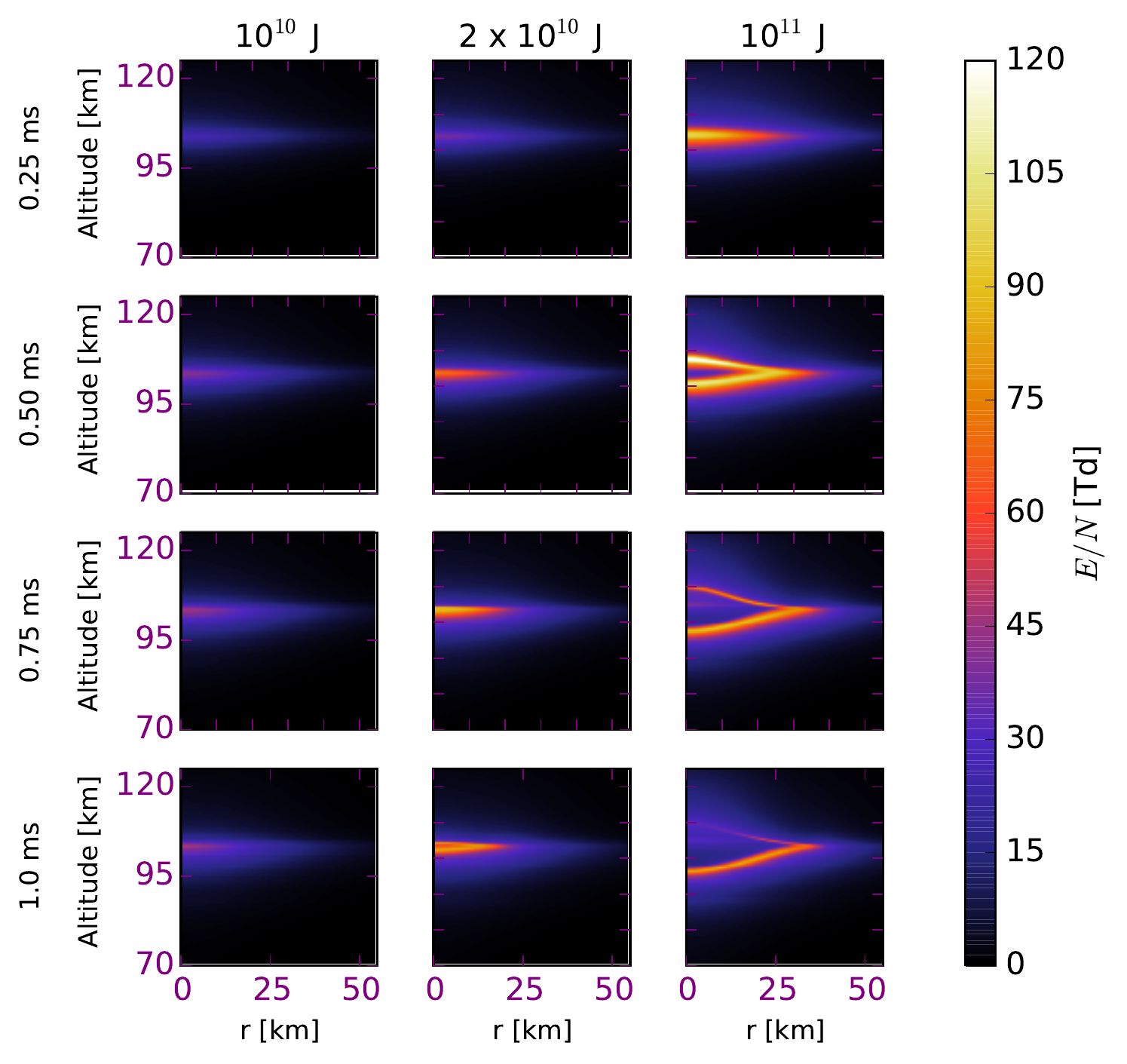}
\caption{\label{Fig5_Venus_nighttime.pdf}  Time evolution of the reduced electric field $E/N$ in the atmosphere of Venus under different IC lightning discharges with total released energy of 10$^{10}$ J, 2 $\times$ 10$^{10}$  J and 10$^{11}$ J. Four snapshots are shown for each energy. The discharge length of the IC channel in this case is 10 km, and the radius of the spheres (acting as clouds) that contain the charges is set to 2.5 km.}
\end{figure}

\begin{figure}
\includegraphics[width=.9\columnwidth] {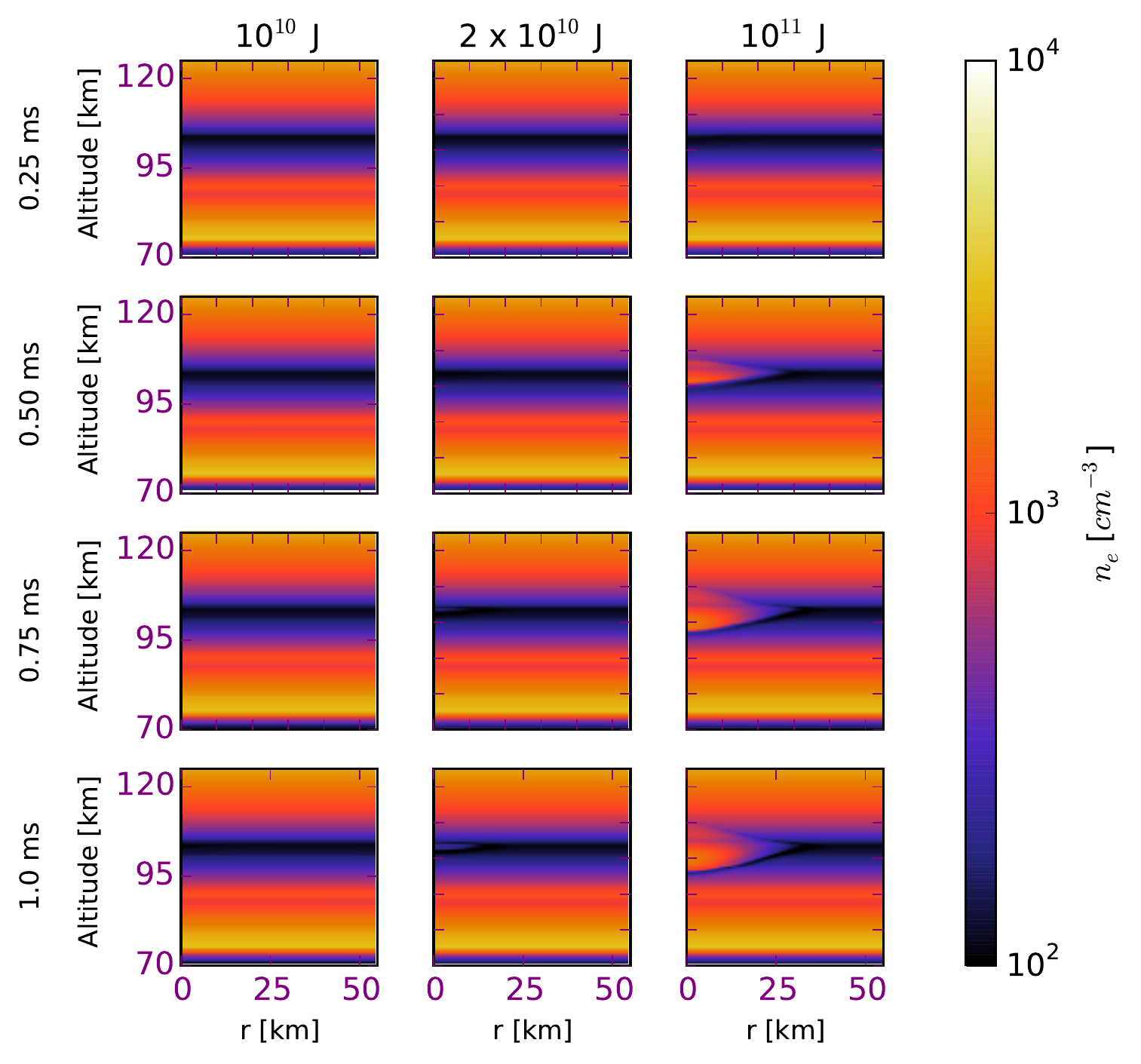}
\caption{\label{Fig6_Venus_nighttime.pdf}  Time evolution of the electron density in the atmosphere of Venus under different IC lightning discharges with total released energy of 10$^{10}$ J, 2 $\times$ 10$^{10}$ J and 10$^{11}$ J. Four snapshots are shown for each energy.}
\end{figure}

\begin{figure}
\includegraphics[width=.9\columnwidth] {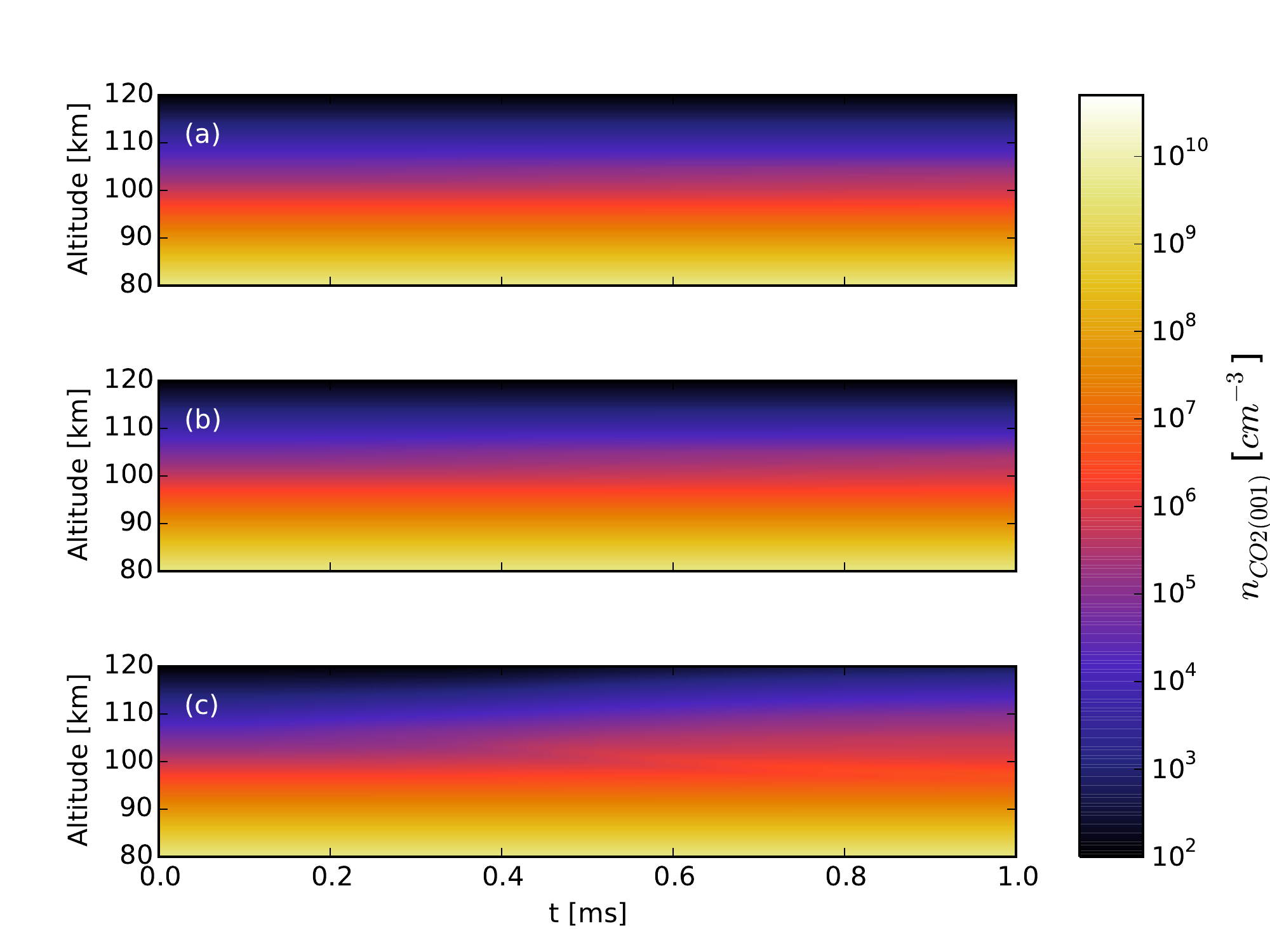}
\caption{\label{Fig7_Venus_nighttime.pdf}  Time evolution of the CO$_2$(001) density in the center of the halo created by IC lightning discharges with total released energy of (a) 10$^{10}$ J, (b) 2 $\times$ 10$^{10}$ J and (c) 10$^{11}$ J.}
\end{figure}

\begin{figure}
\includegraphics[width=.9\columnwidth] {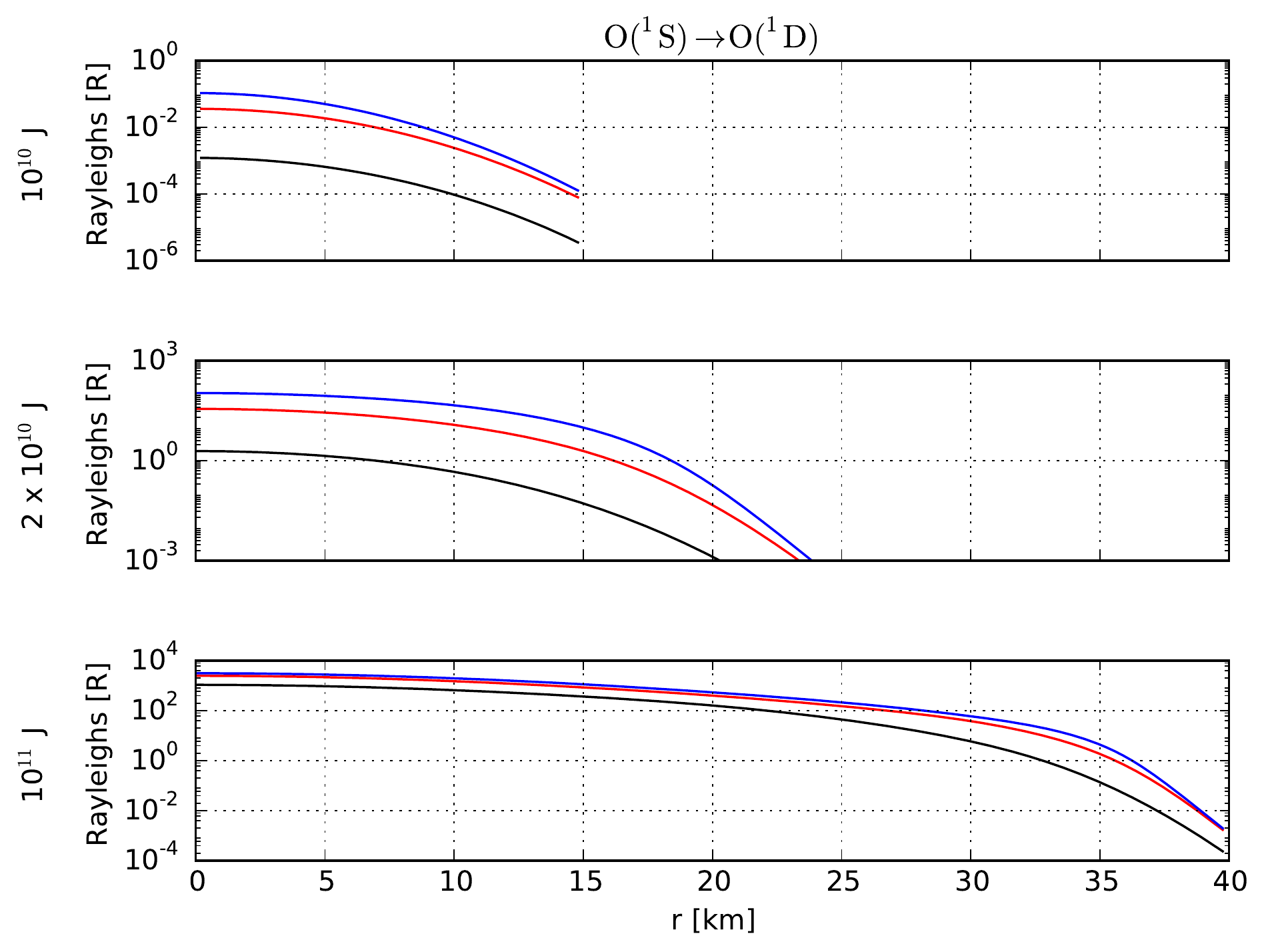}
\caption{\label{Fig8_Venus_nighttime.pdf}  Emission brightness in Rayleighs from the radiative decay  O($^{1}S$)  $\rightarrow$  O($^{1}D$) (557 nm) along different positions of the glowing disc and for different times after the beginning of the IC discharge: 0.5 ms (black line), 0.75 ms (red line) and 1 ms (blue line). The considered IC discharges release a total energy of 10$^{10}$ J (top panel), 2 $\times$ 10$^{10}$ J (middle panel) and 10$^{11}$ J (bottom panel).}
\end{figure}

\begin{figure}
\includegraphics[width=.9\columnwidth] {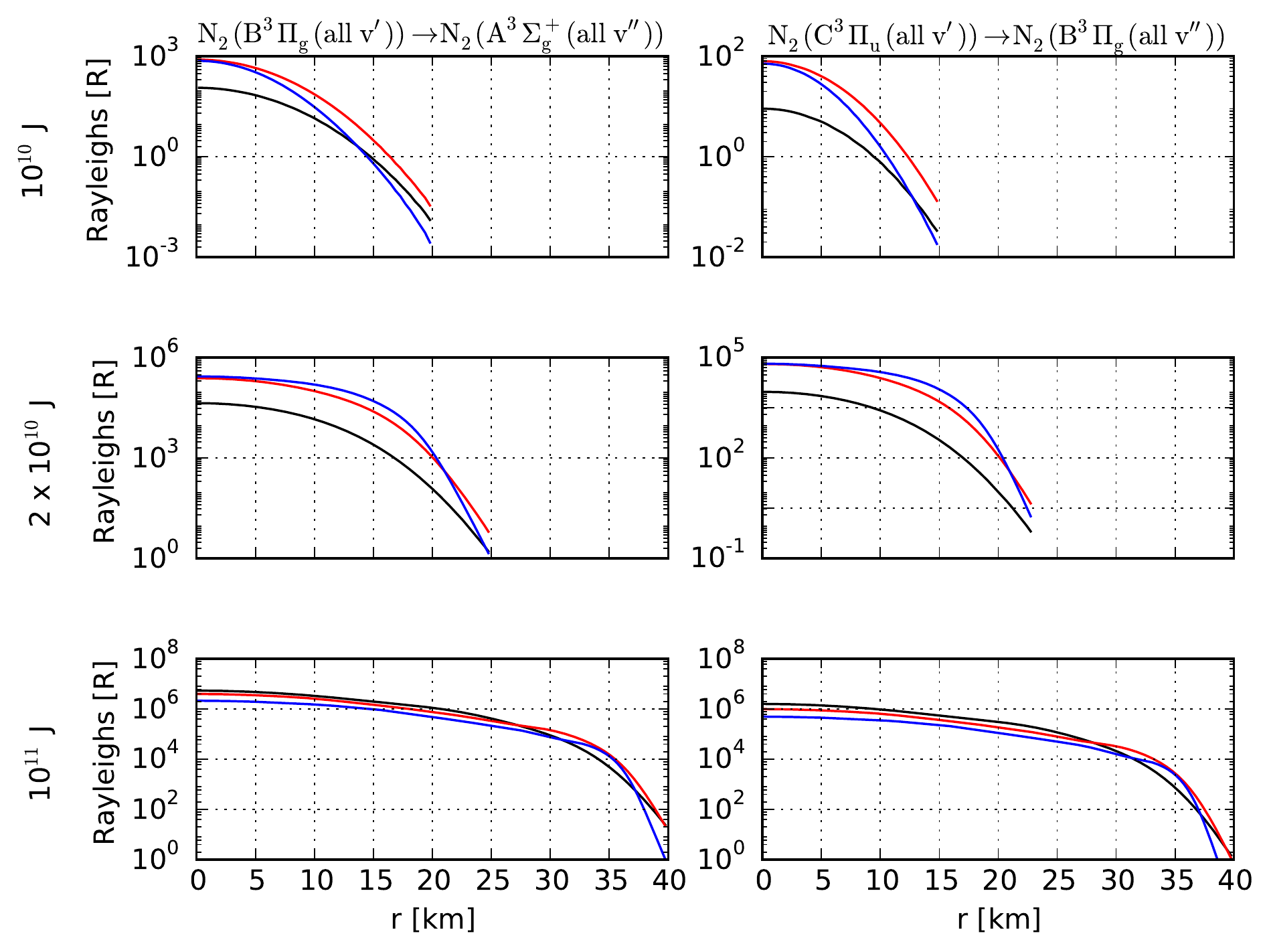}
\caption{\label{Fig9_Venus_nighttime.pdf}  Emission brightness in Rayleighs from the radiative decays N$_2$($B^{3}\Pi_g$ (all v$^{\prime}$)) $\rightarrow$ N$_2$($A^{3}\Sigma_{g}^{+}$ (all v$^{\prime\prime}$)) (550 nm - 1.2 $\mu$m) and N$_2$($C^{3}\Pi_u$ (all v$^{\prime}$)) $\rightarrow$ N$_2$($B^{3}\Pi_g$ (all v$^{\prime\prime}$)) (250 - 450 nm) along different positions of the glowing disc and for different times after the beginning of the IC discharge: 0.5 ms (black line), 0.75 ms (red line) and 1 ms (blue line). The considered IC discharges release a total energy of 10$^{10}$ J (top panel), 2 $\times$ 10$^{10}$ J (middle panel) and 10$^{11}$ J (bottom panel).}
\end{figure}

\begin{figure}
\includegraphics[width=.9\columnwidth] {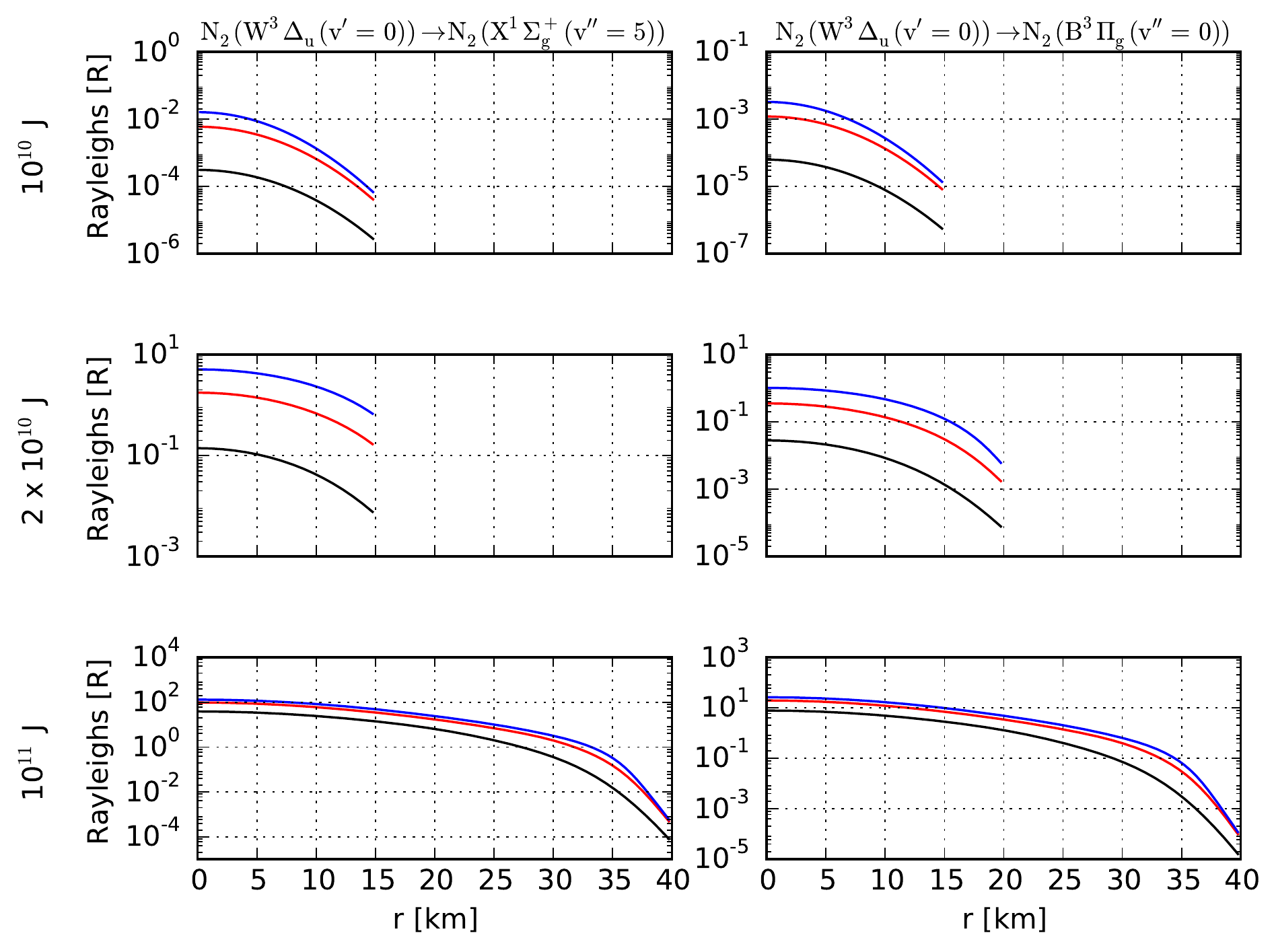}
\caption{\label{Fig10_Venus_nighttime.pdf}  Emission brightness in Rayleighs from the radiative decay N$_2$($W^{3}\Delta_u$ (v$^{\prime}$ = 0))  $\rightarrow$ N$_2$($X^{1}\Sigma_{g}^{+}$ (v$^{\prime\prime}$ = 5)) (208 nm) and N$_2$($W^{3}\Delta_u$ (v$^{\prime}$ = 0))  $\rightarrow$ N$_2$($B^{3}\Pi_g$ (v$^{\prime\prime}$ = 0))  (136.10 $\mu$m) along different position of the glowing disc and for different times after the beginning of the IC discharge: 0.5 ms (black line), 0.75 ms (red line) and 1 ms (blue line). The considered IC discharges release a total energy of 10$^{10}$ J (top panel), 2 $\times$ 10$^{10}$ J (middle panel) and 10$^{11}$ J (bottom panel).}
\end{figure}

\begin{figure}
\includegraphics[width=.9\columnwidth] {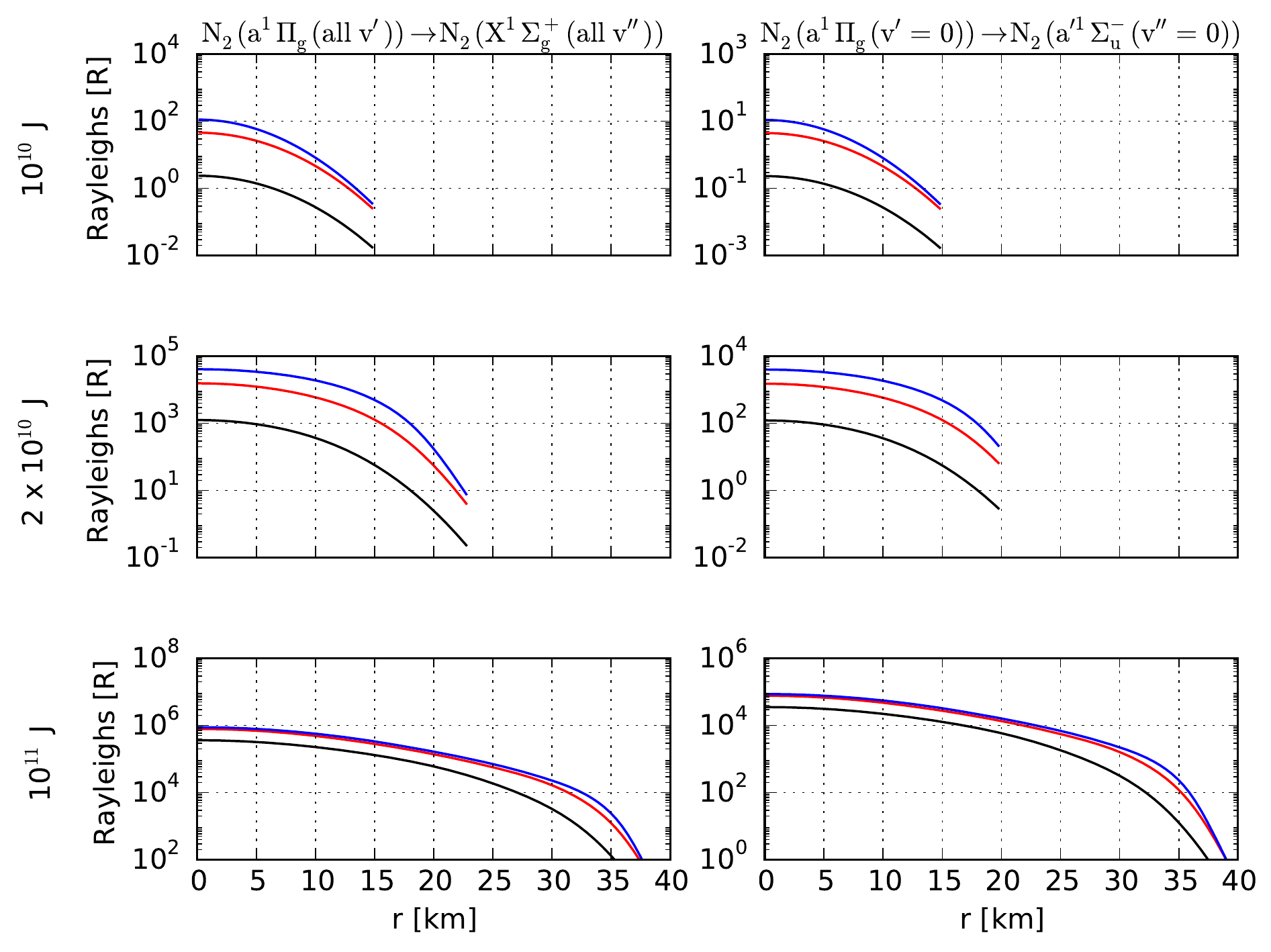}
\caption{\label{Fig11_Venus_nighttime.pdf}  Emission brightness in Rayleighs from the radiative decays N$_2$($a^{1}\Pi_g$ (all v$^{\prime}$))  $\rightarrow$ N$_2$($X^{1}\Sigma_{g}^{+}$ (all v$^{\prime\prime}$)) (120 - 280 nm) and  N$_2$($a^{1}\Pi_g$)  $\rightarrow$ N$_2$($a^{\prime1}\Pi_g$) (8.25 $\mu$m) along different positions of the glowing disc and for different times after the beginning of the IC discharge: 0.5 ms (black line), 0.75 ms (red line) and 1 ms (blue line). The considered IC discharges release a total energy of 10$^{10}$ J (top panel), 2 $\times$ 10$^{10}$ J (middle panel) and 10$^{11}$ J (bottom panel).}
\end{figure}

\begin{figure}
\includegraphics[width=.9\columnwidth] {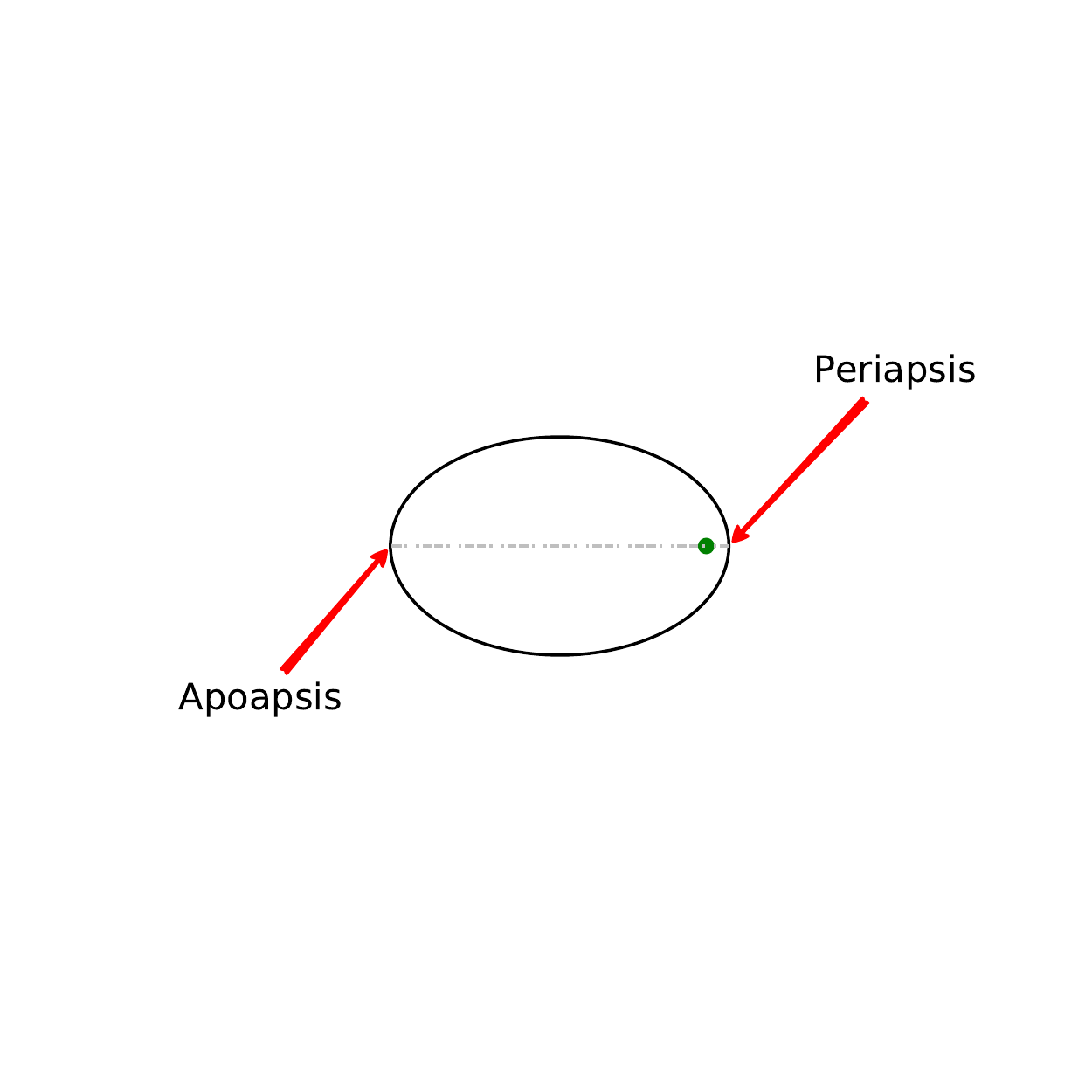}
\caption{\label{Fig12_Venus_nighttime.pdf}  Artistic representation of Akatsuki probe orbiting Venus. The periapsis and apoapsis altitudes after April 2016 are expected to be 300-4000 km and 370000 km respectively \citep{Gibney/Nat} .}
\end{figure}

\begin{figure}
\includegraphics[width=.9\columnwidth] {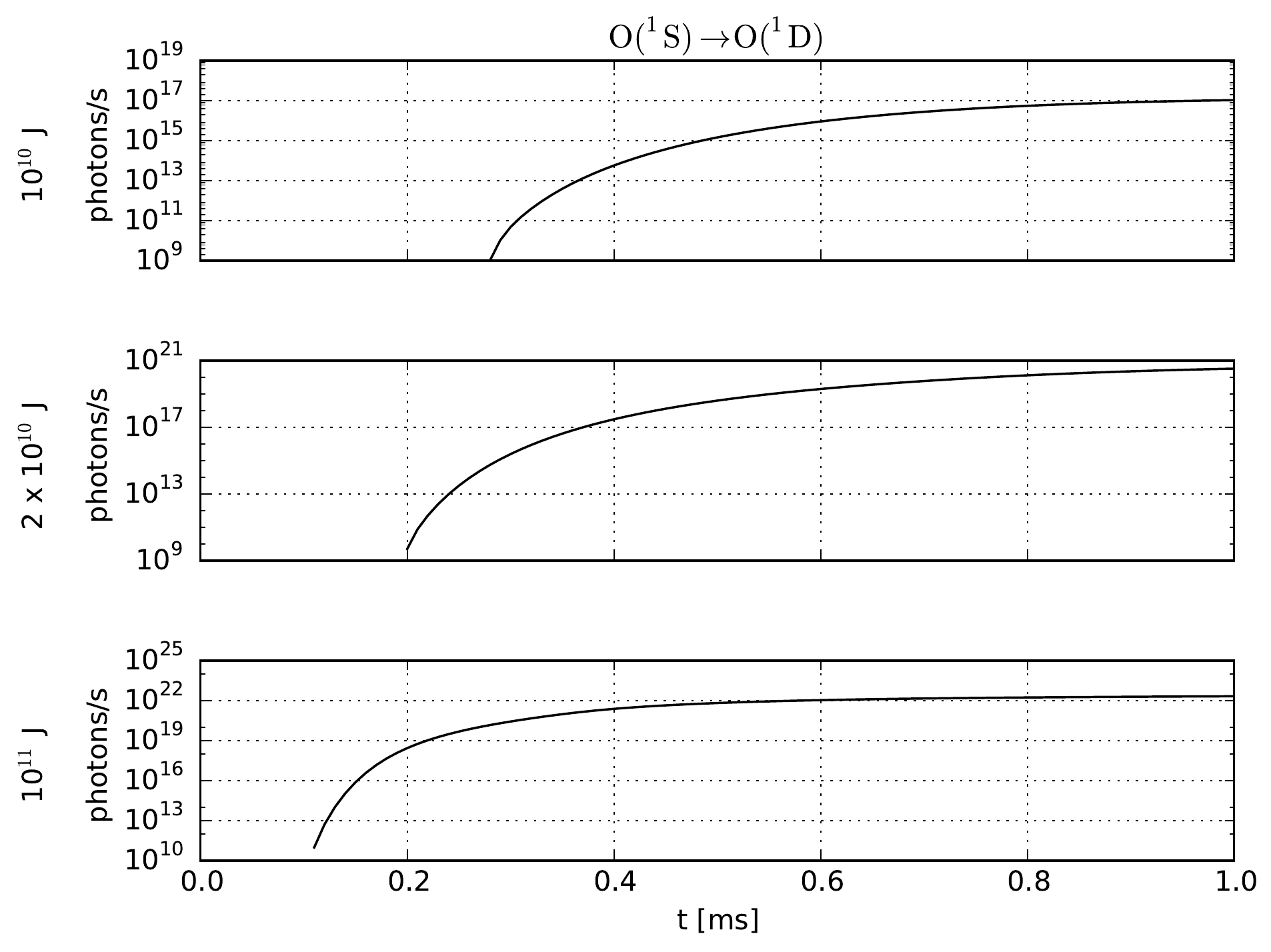}
\caption{\label{Fig13_Venus_nighttime.pdf}  Total number of photons per second emitted by the entire halo up to 1 ms due to the radiative decay 
O($^{1}S$) $\rightarrow$ O + $h\nu$ (557 nm). The considered IC discharges release a total energy of 10$^{10}$ J (top panel), 2 $\times$ 10$^{10}$ J (middle panel) and 10$^{11}$ J (bottom panel).}
\end{figure}

\begin{figure}
\includegraphics[width=.9\columnwidth] {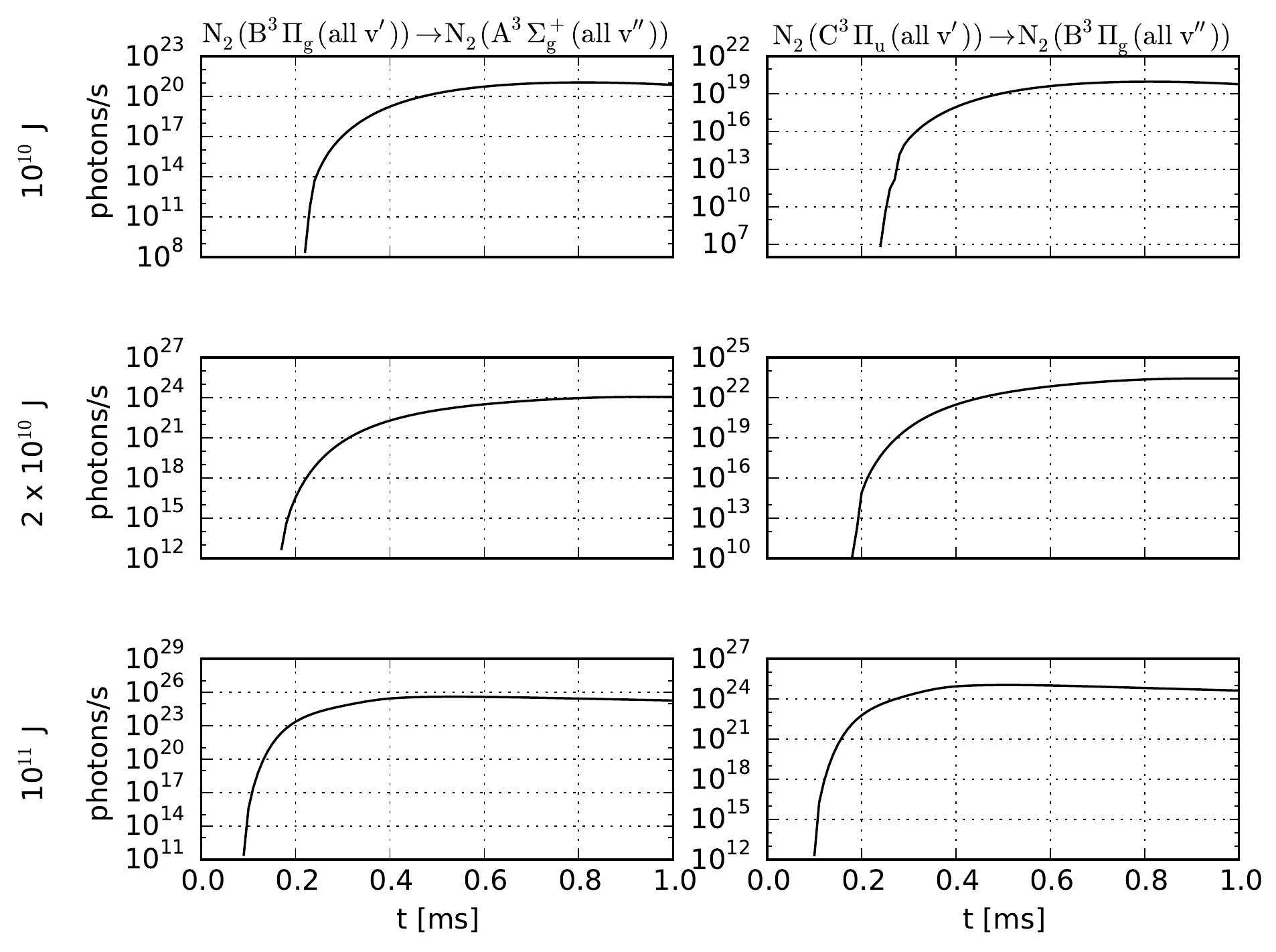}
\caption{\label{Fig14_Venus_nighttime.pdf}  Total number of photons per second emitted by the entire halo up to 1 ms due to the radiative decays  N$_2$($B^{3}\Pi_g$ (all v$^{\prime}$)) $\rightarrow$ N$_2$($A^{3}\Sigma_{g}^{+}$ (all v$^{\prime\prime}$)) + $h\nu$ (550 nm - 1.2 $\mu$m) and N$_2$($C^{3}\Pi_u$ (all v$^{\prime}$)) $\rightarrow$ N$_2$($B^{3}\Pi_g$ (all v$^{\prime\prime}$)) + $h\nu$ (250 - 450 nm). The considered IC discharges release a total energy of 10$^{10}$ J (top panel), 2 $\times$ 10$^{10}$ J (middle panel) and 10$^{11}$ J (bottom panel).}
\end{figure}

\begin{figure}
\includegraphics[width=.9\columnwidth] {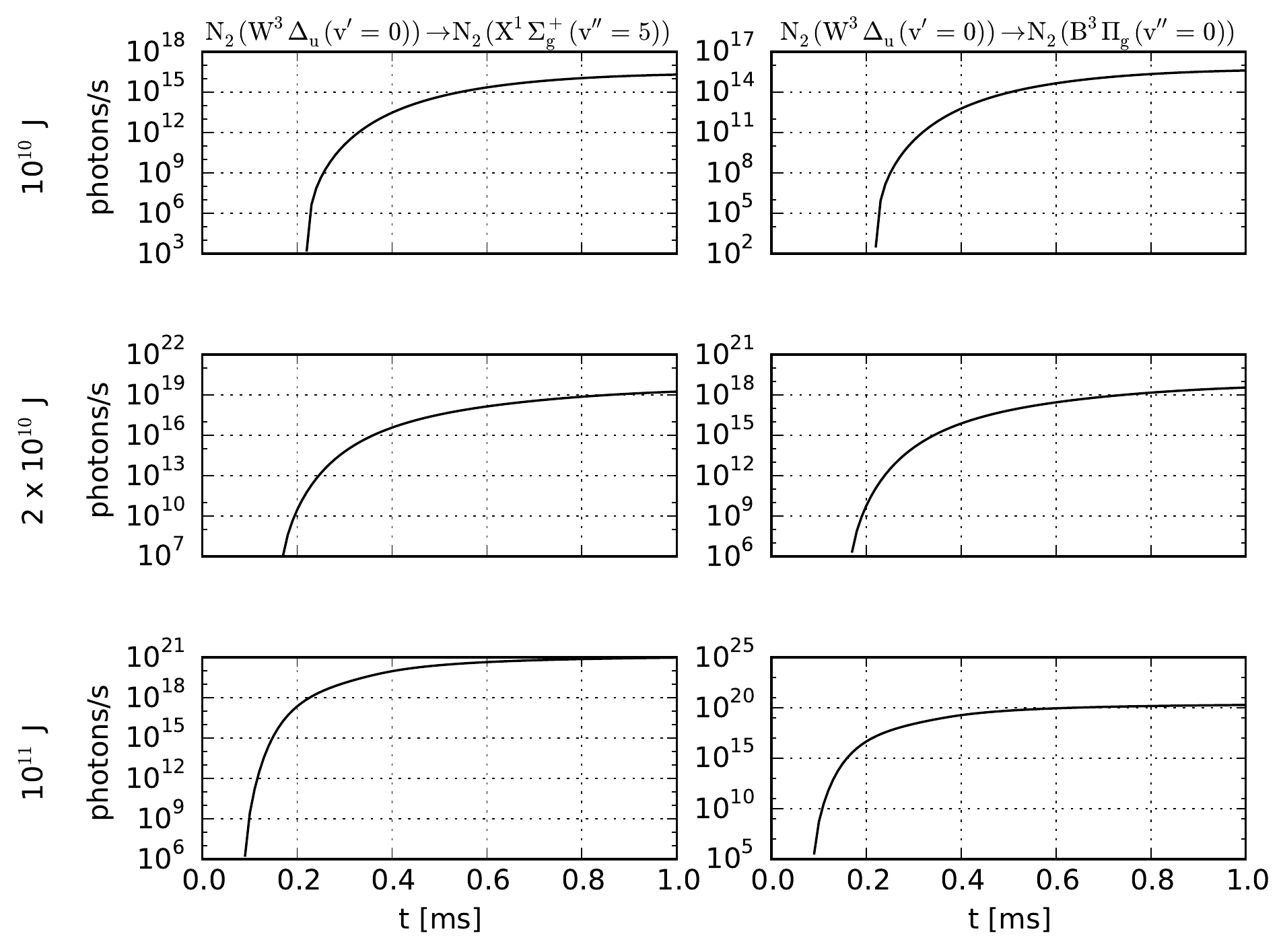}
\caption{\label{Fig15_Venus_nighttime.pdf}  Total number of photons per second emitted by the entire halo up to 1 ms due to the radiative decays N$_2$($W^{3}\Delta_u$ (v$^{\prime}$ = 0))  $\rightarrow$ N$_2$($X^{1}\Sigma_{g}^{+}$ (v$^{\prime\prime}$ = 5)) + $h\nu$ (208 nm)  and N$_2$($W^{3}\Delta_u$ (v$^{\prime}$ = 0))  $\rightarrow$ N$_2$($B^{3}\Pi_g$ (v$^{\prime\prime}$ = 0)) + $h\nu$ (136.10 $\mu$m). The considered IC discharges release a total energy of 10$^{10}$ J (top panel), 2 $\times$ 10$^{10}$ J (middle panel) and 10$^{11}$ J (bottom panel).}
\end{figure}

\begin{figure}
\includegraphics[width=.9\columnwidth] {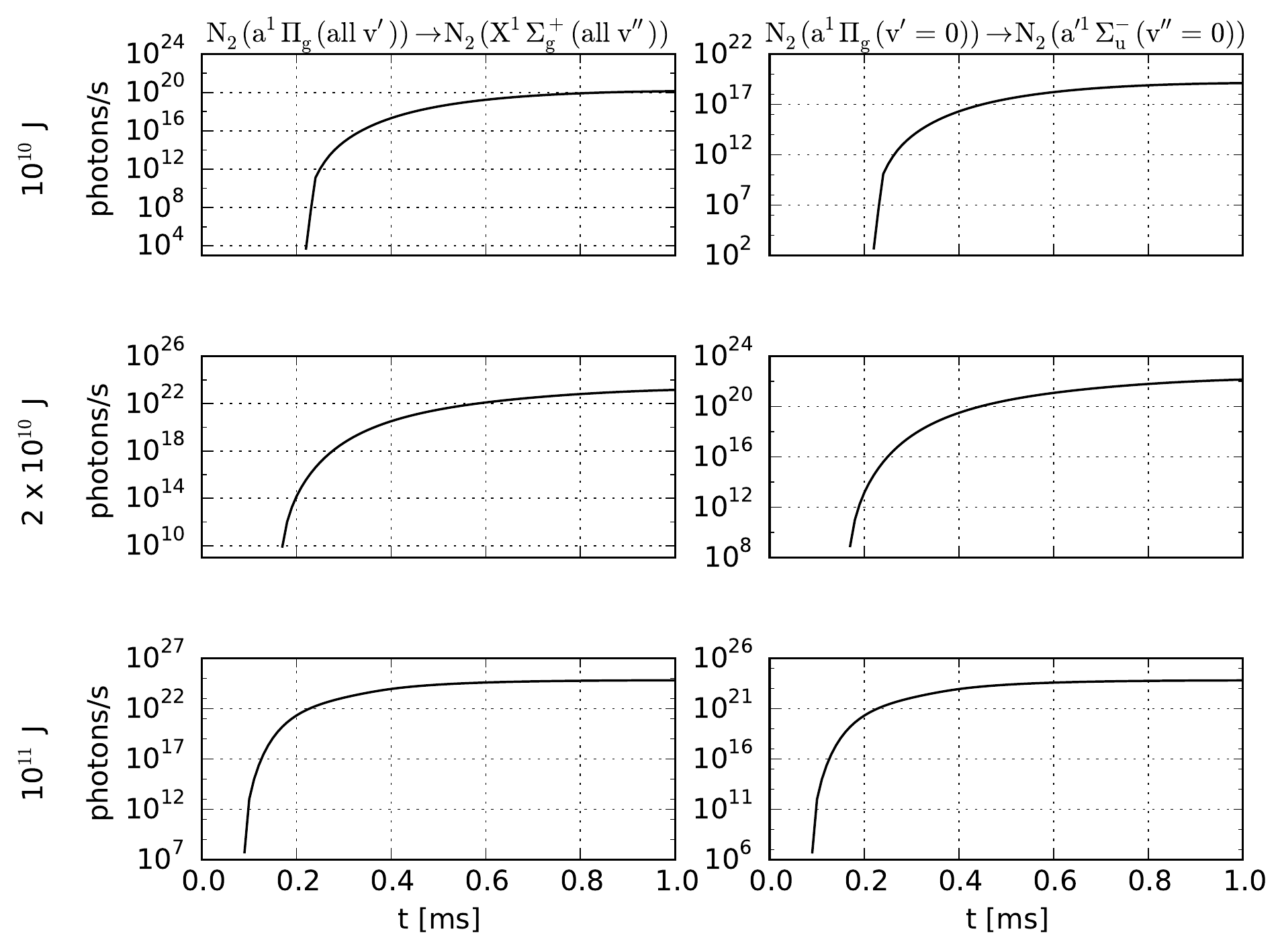}
\caption{\label{Fig16_Venus_nighttime.pdf}  Total number of photons per second emitted by the entire halo up to 1 ms due to the radiative decays N$_2$($a^{1}\Pi_g$ (all v$^{\prime}$))  $\rightarrow$ N$_2$($X^{1}\Sigma_{g}^{+}$ (all v$^{\prime\prime}$)) + $h\nu$ (120 - 280 nm) and  N$_2$($a^{1}\Pi_g$(v$^{\prime}$ = 0))  $\rightarrow$ N$_2$($a^{\prime1}\Pi_g$(v$^{\prime\prime}$ = 0)) + $h\nu$ (8.25 $\mu$m). The considered IC discharges release a total energy of 10$^{10}$ J (top panel), 2 $\times$ 10$^{10}$ J (middle panel) and 10$^{11}$ J (bottom panel).}
\end{figure}

\begin{longtable}{|c|c|}
\caption{Species considered in the basic kinetic model providing ambient nighttime electron density profiles in Venus between 70 km and 125 km of altitude.} \label{tab:long}  \\
\endlastfoot
\hline
     e, O$^-$, CO$_3$$^-$, O$_2$$^-$ \\

     CO$_2$$^+$, O$^+$, O$_2$$^+$, N$_2$$^+$, N$^+$, He$^+$, H$^+$ \\

     H, He, C \\
\hline
\end{longtable}

\begin{longtable}{|c|c|c|c|}

\hline \multicolumn{1}{|c|}{\textbf{Reaction No.}} & \multicolumn{1}{|c|}{\textbf{Reaction}} & \multicolumn{1}{c|}{\textbf{Rate coefficient}} & \multicolumn{1}{c|}{\textbf{Reference}} \\

\endfirsthead

   \multicolumn{1}{c|}{\textbf{Reaction No.}} & \multicolumn{1}{c|}{\textbf{Reaction}} & \multicolumn{1}{c|}{\textbf{Rate coefficient}} & \multicolumn{1}{c}{\textbf{Reference}} \\
\hline
\endhead

\caption{Basic kinetic scheme of the model providing ambient nighttime electron density profiles in Venus between 75 km and 120 km of altitude. The gas temperature (T) is in K. The electron temperature is T$_e$ = T. References are as follows: [1] \cite{Viggiano2005/JCP}; [2] \cite{Gordillo-Vazquez2008/JPhD}; [3] \cite{Brasseur3rd/Book}; [4] \cite{Chen1978/JGR}; [5] \cite{Upadhyay1990/ASR}; [6] \cite{Schunk1980/ROG}; 
[7] \cite{Nordheim2015/Icarus}; [8] \cite{Chen1978/JGR}. Note that for photoionization mechanisms (37)-(40) reactions rates (in cm$^{-3}$ s$^{-1}$) rather than rate coefficients are shown. The values of the reaction rate $k$ and of the nondimensional magnitudes $\eta_i$ = $N_i$/$N$ 
(with $\eta_1$ = 0.9, $\eta_2$ = 0.1, $\eta_3$ = 10$^{-6}$, $\eta_4$ = 10$^{-6}$) are obtained from \cite{Nordheim2015/Icarus} and \cite{Chen1978/JGR}, respectively. } \label{tab:long}  \\
\endlastfoot

\hline
   1 & e + CO$_2$$^+$ $\rightarrow$ CO + O & 6.5 $\times$ 10$^{-7}$ ($\frac{T_e}{300}$)$^{-0.8}$ cm$^{3}$ s$^{-1}$ & [1]   \\

   2 & O$^-$ + CO $\rightarrow$ CO$_2$ + e & 6.72 $\times$ 10$^{-10}$ cm$^{3}$ s$^{-1}$ & [2]   \\

   3 & O$^-$ + CO$_2$ + N$_2$($X^{1}\Sigma_{g}^{+}$) $\rightarrow$ CO$_3$$^-$ + N$_2$($X^{1}\Sigma_{g}^{+}$) & 3.1 $\times$ 10$^{-28}$ ($\frac{T}{300}$)$^{0.5}$ cm$^{6}$ s$^{-1}$ & [3]  \\

   4 & O$^-$ + CO$_2$ + CO$_2$ $\rightarrow$ CO$_3$$^-$ + CO$_2$  & 3.1 $\times$ 10$^{-28}$ ($\frac{T}{300}$)$^{0.5}$ cm$^{6}$ s$^{-1}$ & [3]   \\

   5 & CO$_3$$^-$ + O $\rightarrow$  O$_2$$^-$ +  CO$_2$ & 1.1 $\times$ 10$^{-10}$  $\times$ ($\frac{T}{300}$)$^{0.5}$ cm$^{3}$ s$^{-1}$ & [3]    \\

   6 & O$_2$$^-$ + O $\rightarrow$  O$^-$ +  O$_2$  & 3.3 $\times$ 10$^{-10}$  cm$^{3}$ s$^{-1}$ & [2]    \\

   7 & CO$_2$$^+$ + O $\rightarrow$  CO + O$_2$$^+$  & 1.6 $\times$ 10$^{-10}$ cm$^{3}$ s$^{-1}$  & [4]    \\

   8 & CO$_2$$^+$ + O $\rightarrow$  CO$_2$ + O$^+$  & 9.6 $\times$ 10$^{-11}$ cm$^{3}$ s$^{-1}$  & [4]    \\

   9 & He$^+$ + CO$_2$ $\rightarrow$ CO + O$^+$ + He  & 1.0 $\times$ 10$^{-10}$ cm$^{3}$ s$^{-1}$  & [5]    \\

   10 & He$^+$ + CO$_2$ $\rightarrow$ CO$^+$ + O + He  & 8.7 $\times$ 10$^{-10}$ cm$^{3}$ s$^{-1}$  & [5]    \\

   11 & CO$^+$ + O $\rightarrow$ CO + O$^+$  & 1.4 $\times$ 10$^{-10}$  cm$^{3}$ s$^{-1}$  & [6]    \\

   12 & O$^+$ + H $\rightarrow$ H$^+$ + O  & 2.5 $\times$ 10$^{-11}$ $\times$ T$^{0.5}$ cm$^{3}$ s$^{-1}$  & [6]    \\

   13 & H$^+$ + O $\rightarrow$ O$^+$ + H  & 3.8 $\times$ 10$^{-10}$ cm$^{3}$ s$^{-1}$  & [4]   \\

   14 & O$^+$ + CO$_2$ $\rightarrow$  CO + O$_2$$^+$ &  9.4 $\times$ 10$^{-10}$ cm$^{3}$ s$^{-1}$ & [6]      \\

    15 & He$^+$ + O $\rightarrow$  He + O$^+$ &  5.0 $\times$ 10$^{-11}$ cm$^{3}$ s$^{-1}$ & [4]     \\

    16 & O$_2$$^+$ + e $\rightarrow$  O + O &  8.75 $\times$ 10$^{-6}$ T$_e$$^{\frac{-2}{3}}$ cm$^{3}$ s$^{-1}$ & [4]     \\

    17 & CO$^+$ + e $\rightarrow$  C + O &  1.3 $\times$ 10$^{-5}$ T$_e$$^{0.5}$ cm$^{3}$ s$^{-1}$ & [6]     \\

    18 & CO$^+$ + CO$_2$ $\rightarrow$  CO$_2$$^+$ + CO &  1.0 $\times$ 10$^{-9}$ cm$^{3}$ s$^{-1}$ & [6]     \\

    19 & N$_2$$^+$ + CO$_2$ $\rightarrow$  N$_2$ + CO$_2$$^+$ &  7.7 $\times$ 10$^{-10}$ cm$^{3}$ s$^{-1}$ & [6]     \\

    20 & N$_2$$^+$ + CO $\rightarrow$ CO$^+$ + N$_2$($X^{1}\Sigma_{g}^{+}$) &  7.4 $\times$ 10$^{-11}$ cm$^{3}$ s$^{-1}$ & [6]     \\

    21 & N$_2$$^+$ + O $\rightarrow$ O$^+$ +  N$_2$($X^{1}\Sigma_{g}^{+}$) & 3.6 $\times$ 10$^{-12}$ ($\frac{T_e}{300}$) cm$^{3}$ s$^{-1}$ & [6]    \\

    22 & N$^+$ + CO $\rightarrow$ CO$^+$ + N &  4.0 $\times$ 10$^{-10}$ cm$^{3}$ s$^{-1}$ & [6]     \\

    23 & N$^+$ + CO$_2$ $\rightarrow$ CO$_2$$^+$ + N &  7.5 $\times$ 10$^{-10}$ cm$^{3}$ s$^{-1}$ & [6]     \\

    24 & O$^+$ + e $\rightarrow$ O + $h\nu$ &  3.4 $\times$ 10$^{-12}$ ($\frac{T_e}{300}$)$^{0.5}$ cm$^{3}$ s$^{-1}$ & [5]     \\

    25 & CO$^+$ + e $\rightarrow$ CO &  6.5 $\times$ 10$^{-7}$ ($\frac{T_e}{300}$)$^{-0.53}$ cm$^{3}$ s$^{-1}$ & [5]     \\

    26 & N$^+$ + e $\rightarrow$ N + $h\nu$ &  3.3 $\times$ 10$^{-12}$ ($\frac{T_e}{300}$)$^{-0.5}$ cm$^{3}$ s$^{-1}$ & [5]     \\

    27 & N$_2$$^+$ + e $\rightarrow$ N + N &  2.5 $\times$ 10$^{-7}$ ($\frac{T_e}{300}$)$^{-0.5}$ cm$^{3}$ s$^{-1}$ & [5]     \\

    28 & H$^+$ + e $\rightarrow$ H + $h\nu$ &  4.4 $\times$ 10$^{-12}$ ($\frac{T_e}{300}$)$^{-0.5}$ cm$^{3}$ s$^{-1}$ &  [5]    \\

    29 &  He$^+$ + e $\rightarrow$ He + $h\nu$ &  4.4 $\times$ 10$^{-12}$ ($\frac{T_e}{300}$)$^{-0.5}$ cm$^{3}$ s$^{-1}$ & [5]      \\

    30 & He$^+$ + CO$_2$ $\rightarrow$ CO$^+$ + O + He &  8.7 $\times$ 10$^{-10}$ cm$^{3}$ s$^{-1}$ & [6]     \\

    31 & He$^+$ + CO$_2$ $\rightarrow$ CO$_2$$^+$ + He &  1.2 $\times$ 10$^{-10}$ cm$^{3}$ s$^{-1}$ & [6]     \\

    32 & He$^+$ + CO$_2$ $\rightarrow$ O$^+$ + CO + He &  1.0 $\times$ 10$^{-10}$ cm$^{3}$ s$^{-1}$ & [6]     \\

    33 & He$^+$ + CO $\rightarrow$ CO$^+$ + He &  1.68 $\times$ 10$^{-9}$ cm$^{3}$ s$^{-1}$ & [6]     \\

    34 & He$^+$ + N$_2$($X^{1}\Sigma_{g}^{+}$) $\rightarrow$ N$^+$ + N + He &  9.6 $\times$ 10$^{-10}$ cm$^{3}$ s$^{-1}$ & [6]     \\

    35 & He$^+$ + N$_2$($X^{1}\Sigma_{g}^{+}$) $\rightarrow$ N$_2$$^+$ + He &  6.4 $\times$ 10$^{-10}$ cm$^{3}$ s$^{-1}$ & [6]     \\

    36 & CO$_2$$^+$ + H $\rightarrow$ H$^+$ + CO$_2$ &  1.0$\times$ 10$^{-10}$ cm$^{3}$ s$^{-1}$ & [6]     \\
    
    37 & CO$_2$ + h$\nu$ $\rightarrow$ CO$_2^+$ + e &  $k$ $\times$ $\eta_1$ cm$^{-3}$ s$^{-1}$ & [7], [8]     \\
    
    38 & O + h$\nu$ $\rightarrow$ O$^+$ + e &  $k$ $\times$ $\eta_2$ cm$^{-3}$ s$^{-1}$ & [7], [8]     \\

    39 & He + h$\nu$ $\rightarrow$ He$^+$ + e &  $k$ $\times$ $\eta_3$ cm$^{-3}$ s$^{-1}$ & [7], [8]     \\
    
    40 & H + h$\nu$ $\rightarrow$ H$^+$ + e &  $k$ $\times$ $\eta_4$ cm$^{-3}$ s$^{-1}$ & [7], [8]     \\

\hline

\end{longtable}

\begin{longtable}{|c|c|}
\caption{Species considered in the 2D electric discharge model between 70 km and 125 km of altitude.} \label{tab:long}  \\
\endlastfoot
\hline
     e, O$^-$, CO$_3$$^-$, O$_2$$^-$, O$^-$\\

     CO$_2$$^+$, O$_2$$^+$, N$_2$$^+$ \\

     O($^1$S), O($^1$D), O($^3$P), O($^5$P), C \\

     N$_2$(A$^3\Sigma_u^+$), N$_2$(B$^3\Pi_g$ (all v$^{\prime}$)), N$_2$(W$^3\Delta_u$ (all v$^{\prime}$)), N$_2$(C$^3\Pi_u$ (all v$^{\prime}$)) \\

     N$_2$(a$^1\Pi_g$ (all v$^{\prime}$)), N$_2$(a$'$$^1\Sigma_u^-$ (all v$^{\prime}$)) \\

     CO$_2$(00$^0$1),  CO$_2$(10$^0$0), CO$_2$(01$^1$0), CO$_2$(02$^0$2) \\

     CO$_2$(02$^2$2),  CO$_2$(03$^1$0), CO$_2$(03$^3$0), CO$_2$(11$^1$0) \\
\hline
\end{longtable}

\begin{longtable}{|c|c|c|c|}

\hline \multicolumn{1}{|c|}{\textbf{Reaction No.}} & \multicolumn{1}{|c|}{\textbf{Reaction}} & \multicolumn{1}{c|}{\textbf{Rate coefficient}} & \multicolumn{1}{c|}{\textbf{Reference}} \\

\endfirsthead

   \multicolumn{1}{c|}{\textbf{Reaction No.}} & \multicolumn{1}{c|}{\textbf{Reaction}} & \multicolumn{1}{c|}{\textbf{Rate coefficient}} & \multicolumn{1}{c}{\textbf{Reference}} \\
\hline
\endhead

\caption{Basic kinetic scheme of the 2D electric discharge model between 70 km and 125 km of altitude in the mesosphere of Venus. The gas temperature (T) is in K. The electron temperature (T$_e$) dependence of some rate coefficients is transformed into a reduced electric field $(E/N)$ dependence following $T_e$(eV) = 2$\bar{\epsilon}$/3 where the mean electron energy is obtained from BOLSIG+.
References are as follows: [1] \cite{Viggiano2005/JCP}; [2] \cite{Gordillo-Vazquez2008/JPhD}; [3] \cite{Brasseur3rd/Book}; [4] \cite{Chen1978/JGR}; [5] \cite{Upadhyay1990/ASR}; [6] \cite{Schunk1980/ROG}; [9] \cite{Itikawa2002/JPCRD}; [10] \cite{Itikawa2006/JPCRD}; [11] \cite{PhelpsCO2}; [12] \cite{Laher1990/JPCRD}; [13] \cite{Pagnon1995/JPD}; [14] \cite{Erdman1987/JCP}; [15] \cite{Atkinson1972/JCP}; [16] \cite{Filseth1970/JCP}; [17] \cite{Parra2014/JGR}; [18] \cite{Dagdigian1988/CPL}; [19] \cite{Rayment1978/IJMSIP}; [20] \cite{Rapp1965/JCP}; [21] \cite{Capitelli/Book}; [22] \cite{Gilmore1992/JPCRD}. Note that reactions No. 1 through 7 together with reactions 16 and 19 are the same as in Table 2.} \label{tab:long} \\

\endlastfoot

\hline

 1 & e + CO$_2$$^+$ $\rightarrow$ CO + O & 6.5 $\times$ 10$^{-7}$ ($\frac{T_e}{300}$)$^{-0.8}$ cm$^{3}$ s$^{-1}$ & [1]   \\

   2 & O$^-$ + CO $\rightarrow$ CO$_2$ + e & 6.72 $\times$ 10$^{-10}$ cm$^{3}$ s$^{-1}$ & [2]   \\

   3 & O$^-$ + CO$_2$ + N$_2$($X^{1}\Sigma_{g}^{+}$) $\rightarrow$ CO$_3$$^-$ + N$_2$($X^{1}\Sigma_{g}^{+}$) & 3.1 $\times$ 10$^{-28}$ ($\frac{T}{300}$)$^{0.5}$ cm$^{6}$ s$^{-1}$ & [3]  \\

   4 & O$^-$ + CO$_2$ + CO$_2$ $\rightarrow$ CO$_3$$^-$ + CO$_2$  & 3.1 $\times$ 10$^{-28}$ ($\frac{T}{300}$)$^{0.5}$ cm$^{6}$ s$^{-1}$ & [3]   \\

   5 & CO$_3$$^-$ + O $\rightarrow$  O$_2$$^-$ +  CO$_2$ & 1.1 $\times$ 10$^{-10}$  $\times$ ($\frac{T}{300}$)$^{0.5}$ cm$^{3}$ s$^{-1}$ & [3]    \\

   6 & O$_2$$^-$ + O $\rightarrow$  O$^-$ +  O$_2$  & 3.3 $\times$ 10$^{-10}$  cm$^{3}$ s$^{-1}$ & [2]    \\

    7 & CO$_2$$^+$ + O $\rightarrow$  CO + O$_2$$^+$  & 1.6 $\times$ 10$^{-10}$ cm$^{3}$ s$^{-1}$  & [4]    \\

   16  & O$_2$$^+$ + e $\rightarrow$  O + O &  8.75 $\times$ 10$^{-6}$ T$_e$$^{\frac{-2}{3}}$ cm$^{3}$ s$^{-1}$ &  [5]    \\

   19 & N$_2$$^+$ + CO$_2$ $\rightarrow$  N$_2$ + CO$_2$$^+$ &  7.7 $\times$ 10$^{-10}$ cm$^{3}$ s$^{-1}$ & [6]     \\

   41 & e + CO$_2$ $\rightarrow$ CO + O$^-$ &  $k_1 (E/N)$  cm$^{3}$s$^{-1}$ &  [9]   \\

   42 & e + CO$_2$ $\rightarrow$ CO$_2$$^{+}$ + 2e &   $k_2 (E/N)$  cm$^{3}$s$^{-1}$ & [9]   \\

   43 & e  + N$_2$($X^{1}\Sigma_{g}^{+}$) $\rightarrow$ N$_2$$^{+}$ + 2e &  $k_3 (E/N)$ cm$^{3}$ s$^{-1}$ &  [10]    \\

   44 & e  + CO $\rightarrow$ O$^{-}$ + C &  $k_4 (E/N)$ cm$^{3}$ s$^{-1}$ & [20]     \\

   45 &  e + CO$_2$  $\rightarrow$ O($^{1}S$) + CO + e  & $k_5 (E/N)$ cm$^{3}$ s$^{-1}$ & [9]    \\

   46 & O$^{-}$ + N$_2$($X^{1}\Sigma_{g}^{+}$) $\rightarrow$ N$_2$O + e &  $k_6 (E/N)$ cm$^{3}$ s$^{-1}$ & [19]    \\

   47 & e + N$_2$($X^{1}\Sigma_{g}^{+}$) $\rightarrow$ N$_2$($B^{3}\Pi_g$) + e  & $k_7 (E/N)$ cm$^{3}$ s$^{-1}$ & [10]    \\

   48 & e + N$_2$($X^{1}\Sigma_{g}^{+}$) $\rightarrow$ N$_2$($C^{3}\Pi_u$) + e  & $k_8 (E/N)$ cm$^{3}$ s$^{-1}$ & [10]    \\

   49 & e + N$_2$($X^{1}\Sigma_{g}^{+}$) $\rightarrow$ N$_2$($W^{3}\Delta_u$) + e  & $k_9 (E/N)$ cm$^{3}$ s$^{-1}$ & [10]    \\

   50 & e + N$_2$($X^{1}\Sigma_{g}^{+}$) $\rightarrow$ N$_2$($a^{1}\Pi_g$) + e  & $k_{10} (E/N)$ cm$^{3}$ s$^{-1}$ & [10]    \\

   51 & e + CO$_2$ $\rightarrow$ CO$_2$(00$^{0}$1) + e  & $k_{11} (E/N)$ cm$^{3}$ s$^{-1}$ & [11]    \\

   52 & e + CO$_2$ $\rightarrow$ CO$_2$(10$^{0}$0) + e  & $k_{12} (E/N)$ cm$^{3}$ s$^{-1}$ & [11]   \\

   53 & e + CO$_2$ $\rightarrow$ CO$_2$(01$^{1}$0) + e  & $k_{13} (E/N)$ cm$^{3}$ s$^{-1}$ & [11]    \\

   54 & e + CO$_2$(00$^{0}$1) $\rightarrow$ CO$_2$ + e  & $k_{14} (E/N)$ cm$^{3}$ s$^{-1}$ & [11]    \\

   55 & e + CO$_2$(10$^{0}$0) $\rightarrow$ CO$_2$ + e  & $k_{15} (E/N)$ cm$^{3}$ s$^{-1}$ & [11]    \\

   56 & e + CO$_2$(01$^{1}$0) $\rightarrow$ CO$_2$ + e  & $k_{16} (E/N)$ cm$^{3}$ s$^{-1}$ & [11]    \\

   57 & e + O $\rightarrow$ O($^{3}P$) + e  & $k_{17} (E/N)$ cm$^{3}$ s$^{-1}$ &   [12]  \\

   58 & e + O $\rightarrow$ O($^{5}P$) + e  & $k_{18} (E/N)$ cm$^{3}$ s$^{-1}$ &  [12]   \\

   59 & e + O$_2$ $\rightarrow$ O($^{3}P$) + O + e  & $k_{19} (E/N)$ cm$^{3}$ s$^{-1}$ &   [13]  \\

   60 & e + O$_2$ $\rightarrow$ O($^{5}P$) + O + e  & $k_{20} (E/N)$ cm$^{3}$ s$^{-1}$ &  [14]   \\

   61 & O($^{1}S$) + CO$_2$ $\rightarrow$ Products  & 4.8 $\times$ 10$^{-14}$ cm$^{3}$ s$^{-1}$ &   [15]     \\

   62 & O($^{1}S$) + O$_2$ $\rightarrow$ Products  & 7.4 $\times$ 10$^{-14}$ cm$^{3}$ s$^{-1}$ &    [15]    \\

   63 & O($^{1}S$) + N$_2$($X^{1}\Sigma_{g}^{+}$) $\rightarrow$ O($^{1}D$) + $h\nu$ & 5.0 $\times$ 10$^{-17}$ cm$^{3}$ s$^{-1}$ &    [15]    \\

   64 & O($^{1}S$) + CO $\rightarrow$ Products  & 4.9 $\times$ 10$^{-15}$ cm$^{3}$ s$^{-1}$ &     [16]   \\

   65 & O($^{1}S$) $\rightarrow$ O($^{1}D$) + $h\nu$  & 1.35 s$^{-1}$ &    [2]    \\

   66 & O($^{1}D$) $\rightarrow$ O + $h\nu$  & 5.1 $\times$ 10$^{-3}$ s$^{-1}$ &    [2]    \\

   67 & O($^{1}D$) + CO$_2$ $\rightarrow$ Products  & 1.1 $\times$ 10$^{-10}$ cm$^{3}$ s$^{-1}$ &    [2]    \\

   68 & O($^{1}D$) + N$_2$($X^{1}\Sigma_{g}^{+}$) $\rightarrow$ Products  & 7.0 $\times$ 10$^{-11}$ cm$^{3}$ s$^{-1}$ &   [2]      \\

   69 & O($^{1}D$) + CO $\rightarrow$ Products  & 8.0 $\times$ 10$^{-11}$ cm$^{3}$ s$^{-1}$ &    [2]     \\

   70 & N$_2$($B^{3}\Pi_g$ (all v$^{\prime}$))  $\rightarrow$ N$_2$($A^{3}\Sigma_{g}^{+}$ (all  v$^{\prime\prime}$)) + $h\nu$  & 1.34 $\times$ 10$^{5}$ s$^{-1}$ &  [21]    \\

   71 & N$_2$($C^{3}\Pi_u$ (all v$^{\prime}$)) $\rightarrow$ N$_2$($B^{3}\Pi_g$ (all v$^{\prime\prime}$) ) + $h\nu$  & 2.45 $\times$ 10$^{7}$ s$^{-1}$ &    [21]    \\

   72 & N$_2$($W^{3}\Delta_u$ (v$^{\prime}=$ 0) )   $\rightarrow$ N$_2$($X^{1}\Sigma_{g}^{+}$ (v$^{\prime\prime}=$ 5))  + $h\nu$  & 0.154 s$^{-1}$  &   [21]    \\
    
   73 & N$_2$($W^{3}\Delta_u$ (v$^{\prime}=$ 0))  $\rightarrow$ N$_2$($B^{3}\Pi_g$ (v$^{\prime\prime}=$ 0) ) + $h\nu$  & 3.1 $\times$ 10$^{-2}$ s$^{-1}$ &   [22]    \\

   74 & N$_2$($a^{1}\Pi_g$ (v$^{\prime}=$ 0)) $\rightarrow$ N$_2$($a^{\prime1}\Sigma_{u}^{-}$ (v$^{\prime\prime}=$ 0) ) + $h\nu$  & 9.74 $\times$ 10$^{1}$ s$^{-1}$ &    [22]    \\

   75&  N$_2$($a^{1}\Pi_g$ (all v$^{\prime}$)  $\rightarrow$ N$_2$($X^{1}\Sigma_{g}^{+}$ (all v$^{\prime\prime}$)) + $h\nu$  & 1.0 $\times$ 10$^{3}$ s$^{-1}$ &    [22]     \\

   76 & CO$_2$(00$^{0}$1) $\rightarrow$ CO$_2$(02$^{0}$0)+ $h\nu$  & 0.2 s$^{-1}$ &    [17]    \\

   77 & CO$_2$(00$^{0}$1) $\rightarrow$ CO$_2$+ $h\nu$  &  450.0 s$^{-1}$ &    [17]   \\

   78 & CO$_2$(10$^{0}$0) $\rightarrow$ CO$_2$(01$^{1}$0)+ $h\nu$  &  2.08 s$^{-1}$ &    [17]   \\

   79 & CO$_2$(01$^{1}$0) $\rightarrow$ CO$_2$+ $h\nu$  &  1.564 s$^{-1}$ &    [17]    \\

   80 & O($^{5}P$) + O$_2$ $\rightarrow$ O+ O$_2$  &  1.08 $\times$ 10$^{-9}$ cm$^{3}$ s$^{-1}$ &   [18]     \\

   81 & O($^{5}P$) + N$_2$($X^{1}\Sigma_{g}^{+}$) $\rightarrow$ O+ N$_2$($X^{1}\Sigma_{g}^{+}$) & 1.08 $\times$ 10$^{-9}$ cm$^{3}$ s$^{-1}$ &   [18]     \\

   82 & O($^{3}P$) + O$_2$ $\rightarrow$ O+ O$_2$  &  7.8 $\times$ 10$^{-10}$ cm$^{3}$ s$^{-1}$ &   [18]     \\

   83 & O($^{3}P$) + N$_2$($X^{1}\Sigma_{g}^{+}$) $\rightarrow$ O+ N$_2$($X^{1}\Sigma_{g}^{+}$)  & 5.9 $\times$ 10$^{-10}$ cm$^{3}$ s$^{-1}$ &   [18]     \\

   84 & O($^{3}P$) + O$_2$ $\rightarrow$ O($^{5}P$)+ O$_2$  &  6 $\times$ 10$^{-11}$ cm$^{3}$ s$^{-1}$ &   [18]     \\

   85 & O($^{3}P$) + N$_2$($X^{1}\Sigma_{g}^{+}$)$\rightarrow$ O($^{5}P$)+ N$_2$($X^{1}\Sigma_{g}^{+}$) & 2 $\times$ 10$^{-11}$ cm$^{3}$ s$^{-1}$ &   [18]     \\

   86 & O($^{3}P$) $\rightarrow$ O+ $h\nu$  & 2.98 $\times$ 10$^{7}$ s$^{-1}$ &  [18]     \\

   87 & O($^{5}P$) $\rightarrow$ O+ $h\nu$  & 2.56 $\times$ 10$^{7}$ s$^{-1}$ &  [18]     \\
\hline

\end{longtable}

\begin{longtable}{|c|c|c|c|c|c|c|}

\hline \multicolumn{1}{|c|}{\textbf{Reaction No.}} & \multicolumn{1}{|c|}{\textbf{Reaction}} & \multicolumn{1}{c|}{\textbf{g}} & \multicolumn{1}{c|}{\textbf{h}} & \multicolumn{1}{c|}{\textbf{i}} & \multicolumn{1}{c|}{\textbf{j}} & \multicolumn{1}{c|}{\textbf{Reference}} \\

\endfirsthead

   \multicolumn{1}{c|}{\textbf{Reaction No.}} & \multicolumn{1}{c|}{\textbf{Reaction}} & \multicolumn{1}{c|}{\textbf{g}} & \multicolumn{1}{c|}{\textbf{h}} & \multicolumn{1}{c|}{\textbf{i}} & \multicolumn{1}{c|}{\textbf{j}} & \multicolumn{1}{c}{\textbf{Reference}} \\
\hline
\endhead
\hline
   88-89 & CO$_2$(00$^{0}$1)  + CO$_2$ $\rightleftharpoons$ CO$_2$(02$^{0}$0) + CO$_2$  & 0.18 & 7.3 $\times$ 10$^{-14}$ & -850.3 &  86523 &   [23]    \\

   90-91 & CO$_2$(00$^{0}$1)  + CO$_2$ $\rightleftharpoons$ CO$_2$(02$^{2}$0) + CO$_2$  & 0.18 & 7.3 $\times$ 10$^{-14}$ & -850.3 &  86523 &     [23]    \\

   92-93 & CO$_2$(00$^{0}$1)  + CO$_2$ $\rightleftharpoons$ CO$_2$(10$^{0}$0) + CO$_2$  & 0.18 & 7.3 $\times$ 10$^{-14}$ & -850.3 &  86523 &     [23]    \\

   94-95 & CO$_2$(00$^{0}$1)  + CO$_2$ $\rightleftharpoons$ CO$_2$(03$^{1}$0) + CO$_2$  & 0.82 & 7.3 $\times$ 10$^{-14}$ & -850.3 &  86523 &     [23]    \\

   96-97 & CO$_2$(00$^{0}$1)  + CO$_2$ $\rightleftharpoons$ CO$_2$(03$^{3}$0) + CO$_2$  & 0.82 & 7.3 $\times$ 10$^{-14}$ & -850.3 &  86523 &     [23]    \\

   98-99 & CO$_2$(00$^{0}$1)  + CO$_2$ $\rightleftharpoons$ CO$_2$(11$^{1}$0) + CO$_2$  & 0.82 & 7.3 $\times$ 10$^{-14}$ & -850.3 &  86523 &     [23]    \\

   100-101& CO$_2$(01$^{1}$0)  + CO$_2$ $\rightleftharpoons$ CO$_2$ + CO$_2$  & 1.0 & 4.2 $\times$ 10$^{-12}$ & -2988 &  303930 &   [24]     \\

   102-103 & CO$_2$(10$^{0}$0)  + CO$_2$ $\rightleftharpoons$ CO$_2$(01$^{1}$0) + CO$_2$  & 2.5 & 4.2 $\times$ 10$^{-12}$ & -2988  &  303930 &     [24]    \\

   104-105 & CO$_2$(10$^{0}$0)  + CO$_2$ $\rightleftharpoons$ CO$_2$(01$^{1}$0) + CO$_2$(01$^{1}$0)  & 1.0 & 2.5 $\times$ 10$^{-11}$ & 0 & 0  &    [25]    \\

   106-107 & CO$_2$(00$^{0}$1)  + CO$_2$ $\rightleftharpoons$ CO$_2$(02$^{0}$0) + CO$_2$(01$^{1}$0)  & 1.0 & 3.6 $\times$ 10$^{-13}$ & -1660 &  176948 &     [23]    \\

   108-109 & CO$_2$(00$^{0}$1)  + CO$_2$ $\rightleftharpoons$ CO$_2$(02$^{2}$0) + CO$_2$(01$^{1}$0)  & 1.0 & 3.6 $\times$ 10$^{-13}$ & -1660 &  176948 &    [23]    \\

   110-111 & CO$_2$(00$^{0}$1)  + CO$_2$ $\rightleftharpoons$ CO$_2$(10$^{0}$0) + CO$_2$(01$^{1}$0)  & 1.0 & 3.6 $\times$ 10$^{-13}$ & -1660 &  176948 &    [23]    \\
\hline
\caption{Vibrational-Translational (VT) and Vibrational-Vibrational (VV) processes. Rates defined as $k_{co2}$ = $g \times h \times exp(i/T + j/T^{2})$ in cm$^{3}$s$^{-1}$ with the gas temperature (T) in Kelvins. The rates of the return processes are calculated multiplying the direct reaction rate by $exp(-E/\kappa_{B}T)$, where $E$ is the energy emitted/absorbed during the process. References are as follows: [23] \cite{Lepoutre1977/CPL}; [24] \cite{Valverde1990/thesis}; [25] \cite{Orr1987/JPC}} \label{tab:long}  \\
\endlastfoot

\hline

\hline

\end{longtable}

\begin{longtable}{|c|c|c|c|}

\hline \multicolumn{1}{|c|}{\textbf{Transition}} & \multicolumn{1}{|c|}{\textbf{Wavelength}} & \multicolumn{1}{|c|}{\textbf{Decay constant (s$^{-1}$)}} & \multicolumn{1}{|c|}{\textbf{Reference}}    \\

\endfirsthead

   \multicolumn{1}{|c|}{\textbf{Transition}} & \multicolumn{1}{|c|}{\textbf{Wavelength}} & \multicolumn{1}{|c|}{\textbf{Decay constant (s$^{-1}$)}} & \multicolumn{1}{|c|}{\textbf{Reference}}       \\
\hline
\endhead
\hline

   O($^{1}S$)  $\rightarrow$  O($^{1}D$) + $h\nu$ & 557 nm & 1.351  & [2] \\

  O($^{1}D$) $\rightarrow$ O + $h\nu$  & 630 nm & 3 $\times$ 10$^{-3}$  & [2] \\

   N$_2$($B^{3}\Pi_g$ (all v$^{\prime}$))  $\rightarrow$ N$_2$($A^{3}\Sigma_{g}^{+}$  (all v$^{\prime\prime}$))+ $h\nu$ & 550 nm - 1.2 $\mu$m & 1.34 $\times$ 10$^{5}$  & [19] \\

  N$_2$($C^{3}\Pi_u$ (all v$^{\prime}$)) $\rightarrow$ N$_2$($B^{3}\Pi_g$ (all v$^{\prime\prime}$))   + $h\nu$ & 250 - 450 nm & 2.45 $\times$ 10$^{7}$  & [19] \\

  N$_2$($W^{3}\Delta_u$)  (v$^{\prime}=$ 0))  $\rightarrow$ N$_2$($X^{1}\Sigma_{g}^{+}$  (v$^{\prime\prime}=$ 5)) + $h\nu$ & 208 nm & 0.154 & [19]  \\

  N$_2$($W^{3}\Delta_u$) (v$^{\prime}=$ 0)) $\rightarrow$ N$_2$($B^{3}\Pi_g$ (v$^{\prime\prime}=$ 0))  + $h\nu$ & 136.10 $\mu$m & 3.11 $\times$ 10$^{-2}$ & [20]  \\

 N$_2$($W^{3}\Delta_u$) (v$^{\prime}=$ 1)) $\rightarrow$ N$_2$($B^{3}\Pi_g$ (v$^{\prime\prime}=$ 0))  + $h\nu$ & 6.4311 $\mu$m & 2.20 $\times$ 10$^{2}$ & [20]  \\

 N$_2$($W^{3}\Delta_u$) (v$^{\prime}=$ 2)) $\rightarrow$ N$_2$($B^{3}\Pi_g$ (v$^{\prime\prime}=$ 0))  + $h\nu$ & 3.3206 $\mu$m & 7.36 $\times$ 10$^{2}$ & [20]  \\

 N$_2$($W^{3}\Delta_u$) (v$^{\prime}=$ 3)) $\rightarrow$ N$_2$($B^{3}\Pi_g$ (v$^{\prime\prime}=$ 0))  + $h\nu$ & 2.2505 $\mu$m & 8.73 $\times$ 10$^{2}$ & [20]  \\

 N$_2$($W^{3}\Delta_u$) (v$^{\prime}=$ 4)) $\rightarrow$ N$_2$($B^{3}\Pi_g$ (v$^{\prime\prime}=$ 0))  + $h\nu$ & 1.7092 $\mu$m & 6.53 $\times$ 10$^{2}$ & [20]  \\

 N$_2$($W^{3}\Delta_u$) (v$^{\prime}=$ 5)) $\rightarrow$ N$_2$($B^{3}\Pi_g$ (v$^{\prime\prime}=$ 0))  + $h\nu$ & 1.3825 $\mu$m & 3.79 $\times$ 10$^{2}$ & [20]  \\

  N$_2$($a^{1}\Pi_g$) (v$^{\prime}=$ 0)) $\rightarrow$ N$_2$($a^{\prime1}\Sigma_{u}^{-}$ (v$^{\prime\prime}=$ 0))  + $h\nu$ & 8.2515 $\mu$m & 9.74 $\times$ 10$^{1}$ & [20]  \\

  N$_2$($a^{1}\Pi_g$) (v$^{\prime}=$ 1)) $\rightarrow$ N$_2$($a^{\prime1}\Sigma_{u}^{-}$ (v$^{\prime\prime}=$ 0))  + $h\nu$  & 3.4743 $\mu$m & 5.98 $\times$ 10$^{2}$ & [20] \\

  N$_2$($a^{1}\Pi_g$) (v$^{\prime}=$ 2)) $\rightarrow$ N$_2$($a^{\prime1}\Sigma_{u}^{-}$ (v$^{\prime\prime}=$ 0))  + $h\nu$ & 2.2140 $\mu$m & 3.52 $\times$ 10$^{2}$ & [20]  \\

  N$_2$($a^{1}\Pi_g$) (v$^{\prime}=$ 3)) $\rightarrow$ N$_2$($a^{\prime1}\Sigma_{u}^{-}$ (v$^{\prime\prime}=$ 0))  + $h\nu$ & 1.6320 $\mu$m & 5.13 $\times$ 10$^{1}$ & [20]  \\

  N$_2$($a^{1}\Pi_g$) (v$^{\prime}=$ 4)) $\rightarrow$ N$_2$($a^{\prime1}\Sigma_{u}^{-}$ (v$^{\prime\prime}=$ 0))  + $h\nu$ & 1.2969 $\mu$m & 1.83 & [20]  \\

  N$_2$($a^{1}\Pi_g$) (v$^{\prime}=$ 5)) $\rightarrow$ N$_2$($a^{\prime1}\Sigma_{u}^{-}$ (v$^{\prime\prime}=$ 0))  + $h\nu$ & 1.0792 $\mu$m &4.5 $\times$ 10$^{-3}$ & [20]  \\

  N$_2$($a^{1}\Pi_g$) (all v$^{\prime}$)) $\rightarrow$ N$_2$($X^{1}\Sigma_{g}^{+}$ (all v$^{\prime\prime}$)) + $h\nu$ & 120 - 280 nm & 1 $\times$ 10$^{3}$ &[19] and [20]  \\
\hline
\caption{Possible optical emissions due to radiative decay of electronically excited species in the mesosphere of Venus that could be caused by lightning-induced glow discharges. References are as follows: [2] \cite{Gordillo-Vazquez2008/JPhD}; [19] \cite{Capitelli/Book}; [20] \cite{Gilmore1992/JPCRD}.} \label{tab:long}  \\
\endlastfoot

\hline

\hline

\end{longtable}

\begin{longtable}{|c|c|c|}

\hline \multicolumn{1}{|c|}{\textbf{Transition}} & \multicolumn{1}{|c|}{\textbf{Wavelength}} & \multicolumn{1}{|c|}{\textbf{Emitted photons}}   \\

\endfirsthead

   \multicolumn{1}{|c|}{\textbf{Transition}} & \multicolumn{1}{|c|}{\textbf{Wavelength}} & \multicolumn{1}{|c|}{\textbf{Emitted photons}}  \\
\hline
\endhead
\hline

   O($^{1}S$)  $\rightarrow$  O($^{1}D$) + $h\nu$ & 557 nm & 9.26  $\times$ 10$^{16}$  \\

   N$_2$($B^{3}\Pi_g$ (all v$^{\prime}$))   $\rightarrow$ N$_2$($A^{3}\Sigma_{g}^{+}$ (all v$^{\prime\prime}$)) + $h\nu$ &  550 nm - 1.2 $\mu$m & 5.26  $\times$ 10$^{17}$ \\

   N$_2$($C^{3}\Pi_u$ (all v$^{\prime}$))   $\rightarrow$ N$_2$($B^{3}\Pi_g$ (all v$^{\prime\prime}$)) + $h\nu$ &  250 - 450 nm & 4.24 $\times$ 10$^{16}$  \\

  N$_2$($W^{3}\Delta_u$ (v$^{\prime}=$ 0))    $\rightarrow$ N$_2$($X^{1}\Sigma_{g}^{+}$ (v$^{\prime\prime}=$ 5)) + $h\nu$ &   208 nm & 1.27 $\times$ 10$^{13}$  \\
  
  N$_2$($W^{3}\Delta_u$ (v$^{\prime}=$ 0))   $\rightarrow$ N$_2$($B^{3}\Pi_g$ (v$^{\prime\prime}=$ 0)) + $h\nu$   &  136.10 $\mu$m & 6.30 $\times$ 10$^{13}$ \\

  N$_2$($a^{1}\Pi_g$ (v$^{\prime}=$ 0))   $\rightarrow$ N$_2$($a^{\prime1}\Sigma_{u}^{-}$ (v$^{\prime\prime}=$ 0)) + $h\nu$   & 8.25 $\mu$m & 1.77  $\times$ 10$^{16}$  \\

  N$_2$($a^{1}\Pi_g$ (all v$^{\prime}$))   $\rightarrow$ N$_2$($X^{1}\Sigma_{g}^{+}$ (all v$^{\prime\prime}$)) + $h\nu$   & 120 - 280 nm & 1.81 $\times$ 10$^{17}$  \\

\hline
\caption{Total number of emitted photons for IC lightnings with total released energy of 10$^{10}$ J. } \label{tab:long}  \\
\endlastfoot

\hline

\hline

\end{longtable}

\begin{longtable}{|c|c|c|c|}

\hline \multicolumn{1}{|c|}{\textbf{Transition}} & \multicolumn{1}{|c|}{\textbf{Wavelength}} & \multicolumn{1}{|c|}{\textbf{300 km}} & \multicolumn{1}{c|}{\textbf{1000 km}}  \\

\endfirsthead

   \multicolumn{1}{|c|}{\textbf{Transition}} & \multicolumn{1}{|c|}{\textbf{Wavelength}} & \multicolumn{1}{|c|}{\textbf{300 km}} & \multicolumn{1}{c|}{\textbf{1000 km}}  \\
\hline
\endhead
\hline

   O($^{1}S$)  $\rightarrow$  O($^{1}D$) + $h\nu$ & 557 nm  &  40 & 4 \\

   N$_2$($B^{3}\Pi_g$ (all v$^{\prime}$))   $\rightarrow$ N$_2$($A^{3}\Sigma_{g}^{+}$ (all v$^{\prime\prime}$)) + $h\nu$ &  550 nm - 1.2 $\mu$m  &  228 & 21 \\

   N$_2$($C^{3}\Pi_u$ (all v$^{\prime}$))   $\rightarrow$ N$_2$($B^{3}\Pi_g$ (all v$^{\prime\prime}$)) + $h\nu$ &  250 - 450 nm  &  18  & 0 \\

  N$_2$($W^{3}\Delta_u$ (v$^{\prime}=$ 0))    $\rightarrow$ N$_2$($X^{1}\Sigma_{g}^{+}$ (v$^{\prime\prime}=$ 5)) + $h\nu$ &   208 nm  &  0  & 0\\
  
  N$_2$($W^{3}\Delta_u$ (v$^{\prime}=$ 0))   $\rightarrow$ N$_2$($B^{3}\Pi_g$ (v$^{\prime\prime}=$ 0)) + $h\nu$   &  136.10 $\mu$m  &  0 & 0 \\

  N$_2$($a^{1}\Pi_g$ (v$^{\prime}=$ 0))   $\rightarrow$ N$_2$($a^{\prime1}\Sigma_{u}^{-}$ (v$^{\prime\prime}=$ 0)) + $h\nu$   & 8.25 $\mu$m  &  8  & 0  \\

  N$_2$($a^{1}\Pi_g$ (all v$^{\prime}$))   $\rightarrow$ N$_2$($X^{1}\Sigma_{g}^{+}$ (all v$^{\prime\prime}$)) + $h\nu$   & 120 - 280 nm  &  79 & 7 \\

\hline
\caption{Total number of received photons for IC lightnings with total released energy of 10$^{10}$ J. The detector is an hypothetical 25 mm diameter camera located at 300 km or 1000 km away (altitude of 400 km and 1100 km) from the emitting point.} \label{tab:long}  \\
\endlastfoot

\hline

\hline

\end{longtable}

\begin{longtable}{|c|c|c|}

\hline \multicolumn{1}{|c|}{\textbf{Transition}} & \multicolumn{1}{|c|}{\textbf{Wavelength}} & \multicolumn{1}{|c|}{\textbf{Emitted photons}}  \\

\endfirsthead

   \multicolumn{1}{|c|}{\textbf{Transition}} & \multicolumn{1}{|c|}{\textbf{Wavelength}} & \multicolumn{1}{|c|}{\textbf{Emitted photons}}   \\
\hline
\endhead
\hline

   O($^{1}S$)  $\rightarrow$  O($^{1}D$) + $h\nu$ & 557 nm & 2.45  $\times$ 10$^{20}$  \\

   N$_2$($B^{3}\Pi_g$ (all v$^{\prime}$))   $\rightarrow$ N$_2$($A^{3}\Sigma_{g}^{+}$ (all v$^{\prime\prime}$)) + $h\nu$ & 550 nm - 1.2 $\mu$m & 3.84 $\times$ 10$^{20}$ \\

   N$_2$($C^{3}\Pi_u$ (all v$^{\prime}$))   $\rightarrow$ N$_2$($B^{3}\Pi_g$ (all v$^{\prime\prime}$)) + $h\nu$   &250 - 450 nm & 9.74 $\times$ 10$^{19}$ \\

  N$_2$($W^{3}\Delta_u$ (v$^{\prime}=$ 0))    $\rightarrow$ N$_2$($X^{1}\Sigma_{g}^{+}$ (v$^{\prime\prime}=$ 5)) + $h\nu$  &  208 nm & 7.10 $\times$ 10$^{18}$  \\
  
  N$_2$($W^{3}\Delta_u$ (v$^{\prime}=$ 0))   $\rightarrow$ N$_2$($B^{3}\Pi_g$ (v$^{\prime\prime}=$ 0)) + $h\nu$   &  136.10 $\mu$m & 1.43 $\times$ 10$^{18}$  \\

  N$_2$($a^{1}\Pi_g$ (v$^{\prime}=$ 0))   $\rightarrow$ N$_2$($a^{\prime1}\Sigma_{u}^{-}$ (v$^{\prime\prime}=$ 0)) + $h\nu$ &  8.25 $\mu$m & 1.06  $\times$ 10$^{19}$  \\

 N$_2$($a^{1}\Pi_g$ (all v$^{\prime}$))   $\rightarrow$ N$_2$($X^{1}\Sigma_{g}^{+}$ (all v$^{\prime\prime}$)) + $h\nu$ &120 - 280 nm & 3.00 $\times$ 10$^{19}$  \\

\hline
\caption{Total number of emitted photons for IC lightnings with total released energy of 2 $\times$ 10$^{10}$ J.} \label{tab:long}  \\
\endlastfoot

\hline

\hline

\end{longtable}

\begin{longtable}{|c|c|c|c|}

\hline \multicolumn{1}{|c|}{\textbf{Transition}} & \multicolumn{1}{|c|}{\textbf{Wavelength}} & \multicolumn{1}{|c|}{\textbf{300 km}} & \multicolumn{1}{c|}{\textbf{1000 km}}  \\

\endfirsthead

   \multicolumn{1}{|c|}{\textbf{Transition}} & \multicolumn{1}{|c|}{\textbf{Wavelength}} & \multicolumn{1}{|c|}{\textbf{300 km}} & \multicolumn{1}{c|}{\textbf{1000 km}}  \\
\hline
\endhead
\hline

   O($^{1}S$)  $\rightarrow$  O($^{1}D$) + $h\nu$ & 557 nm  &  1.07  $\times$ 10$^{5}$ & 9.5 7 $\times$ 10$^{3}$ \\

   N$_2$($B^{3}\Pi_g$ (all v$^{\prime}$))   $\rightarrow$ N$_2$($A^{3}\Sigma_{g}^{+}$ (all v$^{\prime\prime}$)) + $h\nu$ & 550 nm - 1.2 $\mu$m  &  1.67 $\times$ 10$^{5}$ & 1.50 $\times$ 10$^{4}$  \\

   N$_2$($C^{3}\Pi_u$ (all v$^{\prime}$))   $\rightarrow$ N$_2$($B^{3}\Pi_g$ (all v$^{\prime\prime}$)) + $h\nu$   &250 - 450 nm &   4.22  $\times$ 10$^{4}$ & 3.80 $\times$ 10$^{3}$  \\

  N$_2$($W^{3}\Delta_u$ (v$^{\prime}=$ 0))    $\rightarrow$ N$_2$($X^{1}\Sigma_{g}^{+}$ (v$^{\prime\prime}=$ 5)) + $h\nu$  &  208 nm  &  3.08  $\times$ 10$^{3}$ & 2.77 $\times$ 10$^{2}$  \\
  
  N$_2$($W^{3}\Delta_u$ (v$^{\prime}=$ 0))   $\rightarrow$ N$_2$($B^{3}\Pi_g$ (v$^{\prime\prime}=$ 0)) + $h\nu$   &  136.10 $\mu$m  &  6.22 $\times$ 10$^{2}$ &  57 \\

  N$_2$($a^{1}\Pi_g$ (v$^{\prime}=$ 0))   $\rightarrow$ N$_2$($a^{\prime1}\Sigma_{u}^{-}$ (v$^{\prime\prime}=$ 0)) + $h\nu$ &  8.25 $\mu$m  &  4.61 $\times$ 10$^{3}$ & 4.14 $\times$ 10$^{2}$   \\

 N$_2$($a^{1}\Pi_g$ (all v$^{\prime}$))   $\rightarrow$ N$_2$($X^{1}\Sigma_{g}^{+}$ (all v$^{\prime\prime}$)) + $h\nu$ &120 - 280 nm  &  1.30 $\times$ 10$^{4}$ & 1.17 $\times$ 10$^{3}$  \\

\hline
\caption{Total number of received photons for IC lightnings with total released energy of 2 $\times$ 10$^{10}$ J. The detector is an hypothetical 25 mm diameter camera located at 300 km or 1000 km away (altitude of 400 km and 1100 km) from the emitting point.} \label{tab:long}  \\
\endlastfoot

\hline

\hline

\end{longtable}

\begin{longtable}{|c|c|c|}

\hline \multicolumn{1}{|c|}{\textbf{Transition}} & \multicolumn{1}{|c|}{\textbf{Wavelength}} & \multicolumn{1}{|c|}{\textbf{Emitted photons}}   \\

\endfirsthead

   \multicolumn{1}{|c|}{\textbf{Transition}} & \multicolumn{1}{|c|}{\textbf{Wavelength}} & \multicolumn{1}{|c|}{\textbf{Emitted photons}}   \\
\hline
\endhead
\hline

   O($^{1}S$)  $\rightarrow$  O($^{1}D$) + $h\nu$ & 557 nm & 1.57  $\times$ 10$^{22}$  \\

   N$_2$($B^{3}\Pi_g$ (all v$^{\prime}$))   $\rightarrow$ N$_2$($A^{3}\Sigma_{g}^{+}$ (all v$^{\prime\prime}$)) + $h\nu$ &550 nm - 1.2 $\mu$m & 2.05 $\times$ 10$^{22}$  \\

   N$_2$($C^{3}\Pi_u$ (all v$^{\prime}$))   $\rightarrow$ N$_2$($B^{3}\Pi_g$ (all v$^{\prime\prime}$)) + $h\nu$ & 250 - 450 nm & 5.57 $\times$ 10$^{21}$  \\

  N$_2$($W^{3}\Delta_u$ (v$^{\prime}=$ 0))    $\rightarrow$ N$_2$($X^{1}\Sigma_{g}^{+}$ (v$^{\prime\prime}=$ 5))+  $h\nu$ &208 nm & 1.13 $\times$ 10$^{20}$ \\
  
  N$_2$($W^{3}\Delta_u$ (v$^{\prime}=$ 0))   $\rightarrow$ N$_2$($B^{3}\Pi_g$ (v$^{\prime\prime}=$ 0)) + $h\nu$ & 136.10 $\mu$m & 2.32 $\times$ 10$^{19}$ \\

  N$_2$($a^{1}\Pi_g$ (v$^{\prime}=$ 0))   $\rightarrow$ N$_2$($a^{\prime1}\Sigma_{u}^{-}$ (v$^{\prime\prime}=$ 0)) + $h\nu$ &8.25 $\mu$m & 4.01  $\times$ 10$^{20}$   \\

  N$_2$($a^{1}\Pi_g$ (all v$^{\prime}$))   $\rightarrow$ N$_2$($X^{1}\Sigma_{g}^{+}$ (all v$^{\prime\prime}$)) + $h\nu$ &120 - 280 nm & 2.85 $\times$ 10$^{21}$  \\

\hline
\caption{Total  number of emitted photons for IC lightnings with total released energy of 10$^{11}$ J.} \label{tab:long}  \\
\endlastfoot

\hline

\hline

\end{longtable}

\begin{longtable}{|c|c|c|c|}

\hline \multicolumn{1}{|c|}{\textbf{Transition}} & \multicolumn{1}{|c|}{\textbf{Wavelength}} & \multicolumn{1}{|c|}{\textbf{300 km}} & \multicolumn{1}{c|}{\textbf{1000 km}}  \\

\endfirsthead

   \multicolumn{1}{|c|}{\textbf{Transition}} & \multicolumn{1}{|c|}{\textbf{Wavelength}} & \multicolumn{1}{|c|}{\textbf{300 km}} & \multicolumn{1}{c|}{\textbf{1000 km}}  \\
\hline
\endhead
\hline

   O($^{1}S$)  $\rightarrow$  O($^{1}D$) + $h\nu$ & 557 nm &  6.83 $\times$ 10$^{6}$ & 6.13$\times$ 10$^{5}$ \\

   N$_2$($B^{3}\Pi_g$ (all v$^{\prime}$))   $\rightarrow$ N$_2$($A^{3}\Sigma_{g}^{+}$ (all v$^{\prime\prime}$)) + $h\nu$ &550 nm - 1.2 $\mu$m &   8.90 $\times$ 10$^{6}$ & 8.0 $\times$ 10$^{5}$  \\

   N$_2$($C^{3}\Pi_u$ (all v$^{\prime}$))   $\rightarrow$ N$_2$($B^{3}\Pi_g$ (all v$^{\prime\prime}$)) + $h\nu$ & 250 - 450 nm  &  2.42  $\times$ 10$^{6}$ & 2.18 $\times$ 10$^{5}$  \\

  N$_2$($W^{3}\Delta_u$ (v$^{\prime}=$ 0))    $\rightarrow$ N$_2$($X^{1}\Sigma_{g}^{+}$ (v$^{\prime\prime}=$ 5))+  $h\nu$ &208 nm  &  4.92  $\times$ 10$^{4}$ & 4.41 $\times$ 10$^{3}$  \\
  
  N$_2$($W^{3}\Delta_u$ (v$^{\prime}=$ 0))   $\rightarrow$ N$_2$($B^{3}\Pi_g$ (v$^{\prime\prime}=$ 0)) + $h\nu$ & 136.10 $\mu$m  &  1.01 $\times$ 10$^{4}$ & 9.06 $\times$ 10$^{2}$  \\

  N$_2$($a^{1}\Pi_g$ (v$^{\prime}=$ 0))   $\rightarrow$ N$_2$($a^{\prime1}\Sigma_{u}^{-}$ (v$^{\prime\prime}=$ 0)) + $h\nu$ &8.25 $\mu$m  &  1.74 $\times$ 10$^{5}$ & 1.57 $\times$ 10$^{4}$   \\

  N$_2$($a^{1}\Pi_g$ (all v$^{\prime}$))   $\rightarrow$ N$_2$($X^{1}\Sigma_{g}^{+}$ (all v$^{\prime\prime}$)) + $h\nu$ &120 - 280 nm  &  1.24 $\times$ 10$^{6}$ &  1.11 $\times$ 10$^{5}$ \\

\hline
\caption{Total number of received photons for IC lightnings with total released energy of 10$^{11}$ J. The detector is an hypothetical 25 mm diameter camera located at 300 km or 1000 km away (altitude of 400 km and 1100 km) from the emitting point.} \label{tab:long}  \\
\endlastfoot

\hline

\hline

\end{longtable}

\end{document}